\documentclass[showpacs,preprintnumbers,prb,amsmath,amssymb]{revtex4}
\usepackage{graphicx}
\usepackage{bm}

\begin{document}

\title{
Laser ion acceleration by using the dynamic motion of a target
}

\author{Toshimasa~Morita}

\affiliation{Advanced Beam Technology Research Division,
Japan Atomic Energy Agency, 8-1-7 Umemidai, Kizugawa,
Kyoto 619-0215, Japan}


\begin{abstract}
Proton acceleration by using a 620-TW, $18$-J laser pulse of
peak intensity of $5\times 10^{21}$ W/cm$^{2}$
irradiating a disk target is examined using three-dimensional
particle-in-cell simulations.
It is shown that protons are accelerated efficiently to high energy for
a ``light'' material in the first layer of a double-layer target, because
a strongly inhomogeneous expansion of the first layer occurs
by a Coulomb explosion within such a material.
Moreover, a large movement of the first layer for the accelerated protons
is produced by radiation-pressure-dominant acceleration.
A time-varying electric potential produced by this expanding and
moving ion cloud accelerates protons effectively.
In addition,
using the best material for the target,
one can generate a proton beam with an energy of $200$ MeV and
an energy spread of 2$\%$.
\end{abstract}



\maketitle

\section{INTRODUCTION}

Recently, there has been great progress in compact laser systems,
with dramatic improvements in both laser power and peak intensity.
Ion acceleration by laser pulses has proved to be very useful in applications
using compact laser systems.
Laser-driven fast ions are expected to be useful in many applications
such as hadron therapy, \cite{SBK,MVL}
fast ignition for thermonuclear fusion, \cite{ROT,BRM,ATH}
 laser-driven heavy ion colliders, \cite{ESI1,BEE}
and other applications that use the high-energy ions.
Although the achieved proton energy at present is not high enough for some
applications such as hadron therapy, which requires 200-MeV protons,
other methods can be considered for generating higher energy protons.
One simple way is by using a higher power laser.
However,
current power capabilities of compact lasers are insufficient;
moreover, laser power enhancement will result in a cost increase of the
accelerator.
Therefore, it is important to study conditions for generating higher
energy protons with lower laser power and energy
by using some special techniques. \cite{BWP,FVM,Toncian,HAC,YAH,PRK,PPM,HSM}

In this paper,
I show a way to obtain $200$-MeV protons by using a laser pulse
whose intensity is $I_0 \approx 10^{21}$ W/cm$^{2}$,
energy is $\mathcal{E}_{las} \leq 20$ J,
and power is $P \approx 500$ TW.
I use three-dimensional (3D) particle-in-cell (PIC) simulations
 to investigate how high-energy, high-quality protons
can be generated by a several-hundred-terawatt laser.
I study the proton acceleration during the interaction of the laser pulse with
a double-layer target composed of a high-$Z$ atom layer coated
with a hydrogen layer (see Fig. \ref{fig:fig01}).
As suggested in Refs. \onlinecite{SBK} and \onlinecite{DL,SPJ,MEBKY,MEBKY2},
a quasimonoenergetic ion beam can be obtained using targets of this type.
Our aim is to obtain a high-energy ($\mathcal{E} \approx 200$ MeV) and
high-quality ($\Delta \mathcal{E}/\mathcal{E}\leq 2\%$)
proton beam using a relatively moderate power laser.

In the following sections,
I show the dependence of the proton energy
on the material of the first layer
and that the high-energy protons can be generated by optimally combining
a couple of ion acceleration schemes.

\section{ ION ACCELERATION}
\begin{figure}[tbp]
\includegraphics[clip,width=10.0cm, bb=14 7 457 322]{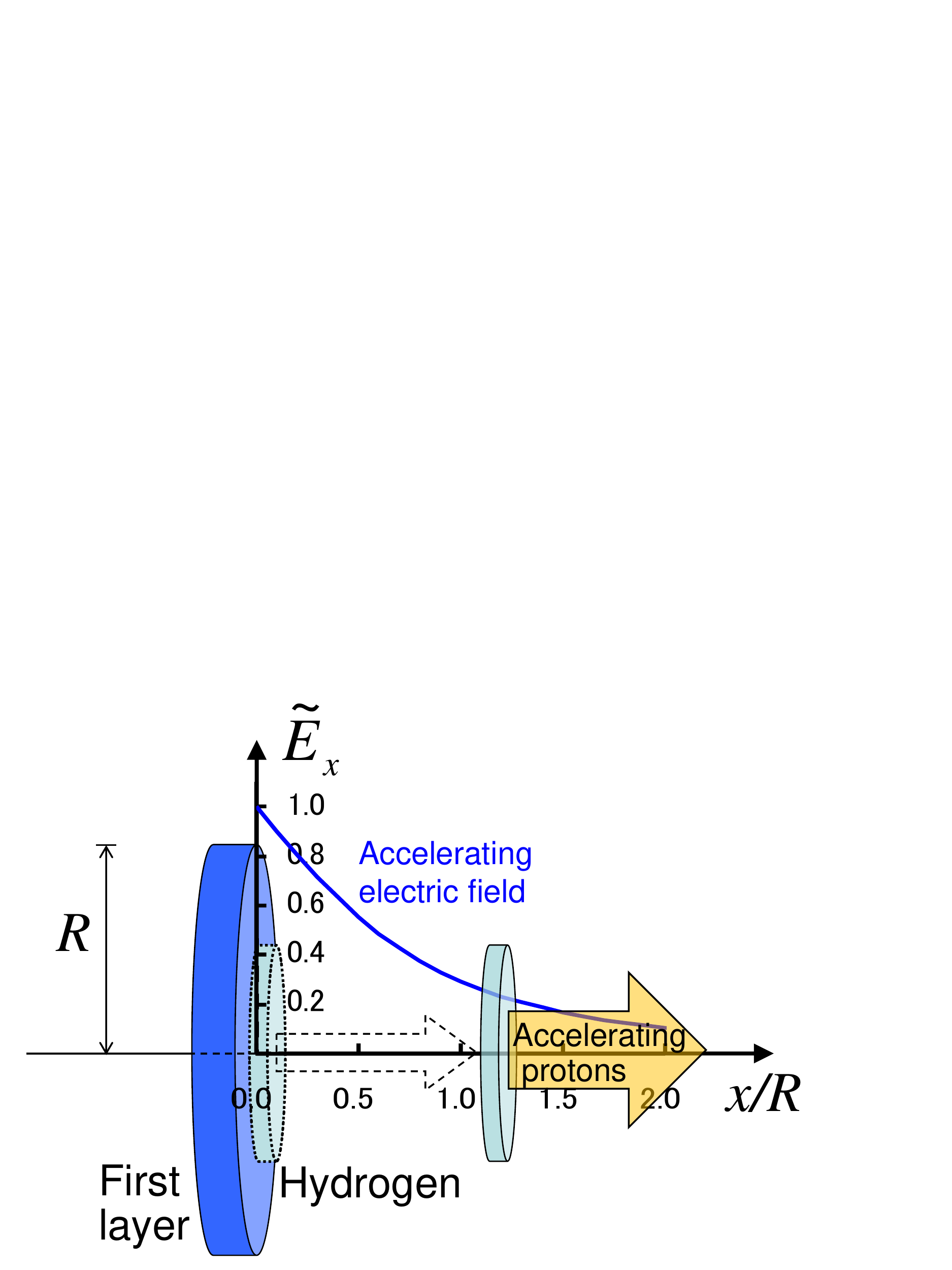}
\caption{
Configuration of a double-layer target. The $x$ component
of the electric field, $\tilde{E}_x(x)$,
is normalized by its maximum, $\rho l/2\epsilon_0$,
of an electrically charged disk on the $x$ axis (solid curve).
Protons are accelerated in this electric field.
}
\label{fig:fig01}
\end{figure}

I consider ion acceleration by a charged disk.
The charged disk is produced by
a laser pulse with sufficiently high intensity irradiating a thin foil.
Many electrons are driven from the foil by the laser pulse,
although the ions of the foils almost stay at their initial positions
because they are much heavier than the electrons.
Therefore,
the thin foil will have a charge, which induces an electrostatic field.
Ions
located on the foil surface are accelerated by this electric field.
The $x$ component of the electric field of a positively charged
thin disk is \begin{equation}
E_x(x)=\frac{\rho l}{2\epsilon_0} \left(1-\frac{x}{\sqrt{x^{2}+R^{2}}} \right),
\label{exx}
\end{equation}
where
$\rho$ is the charge density, $l$ is the disk thickness,
$\epsilon_0$ is the vacuum permittivity,
and $R$ is the charged disk radius.
I assume that the $x$ axis is normal to the disk surface
placed at the disk center.
The solid curve in Fig. \ref{fig:fig01} shows this electric field.
The ions, i.e. protons, are accelerated in this electric field, although
it rapidly decreases as a function of distance from the target surface.
The electric field decreases to $10\%$ at $x =2R$,
which is the distance equal to the diameter of the target and
can be considered to be the spot size of a laser pulse.
Therefore, generating higher energy protons requires producing
a higher surface charge density, $\rho l$, or increasing $R$.
The former requires a higher intensity laser and the latter requires a higher power laser.
The rapidly decreasing accelerating field and its narrow width lead to inefficient proton acceleration by the charged disk.
In this paper,
I present ways to improve these inefficiences
and to generate high-energy protons effectively.

Here, let us define the some terms.
In laser ion acceleration,
the ions are accelerated in some electric field, $E$.
We assume that for an ion of mass of $m$ and charge $q$,
the force on it from the electric field is $qE$.
The equation of motion is $qE=\frac{d}{dt}(mv)$,
where $v$ is the ion velocity.
This equation can be written as
\begin{equation}
E=\frac{d}{dt}(\tilde{m}v),
\label{emv}
\end{equation}
where $\tilde{m}=m/q$.
$\tilde{m}$ is the resistance to movement of an ion
in a certain electric field, $E$;
therefore we call $\tilde{m}$ ``mass'' in this paper.
This expression shows that the smaller $\tilde{m}$ ions can experience greater
acceleration in a certain electric field, $E$.
Therefore, small-``mass'' ions will be called ``light,''
and big-``mass'' ions will be called ``heavy.''
Ions of the same ``mass''
undergo the same movement in a certain electric field.
Note that $\tilde{m}$ is equal to the inverse of the well-known parameter 
$q/m$, the charge-to-mass ratio. I use $\tilde{m}$ in this paper because
it makes it very simple and easy to image the movement of charged particles
in an electric field.

Figure \ref{fig:fig01} shows that
the accelerating protons exit the electric field in a short time.
This means that
the electric field produced is not used enough for proton acceleration.
Therefore, we should create a situation in which
the protons experience this electric field longer
for efficient acceleration.
If the electric potential moves in the direction of the moving protons,
the protons will experience the electric field longer.
In other words, the charged first layer keeps pushing the moving protons.
I present two ways to create this situation.

\begin{figure}[tbp]
\includegraphics[clip,width=10.0cm, bb=12 11 530 375]{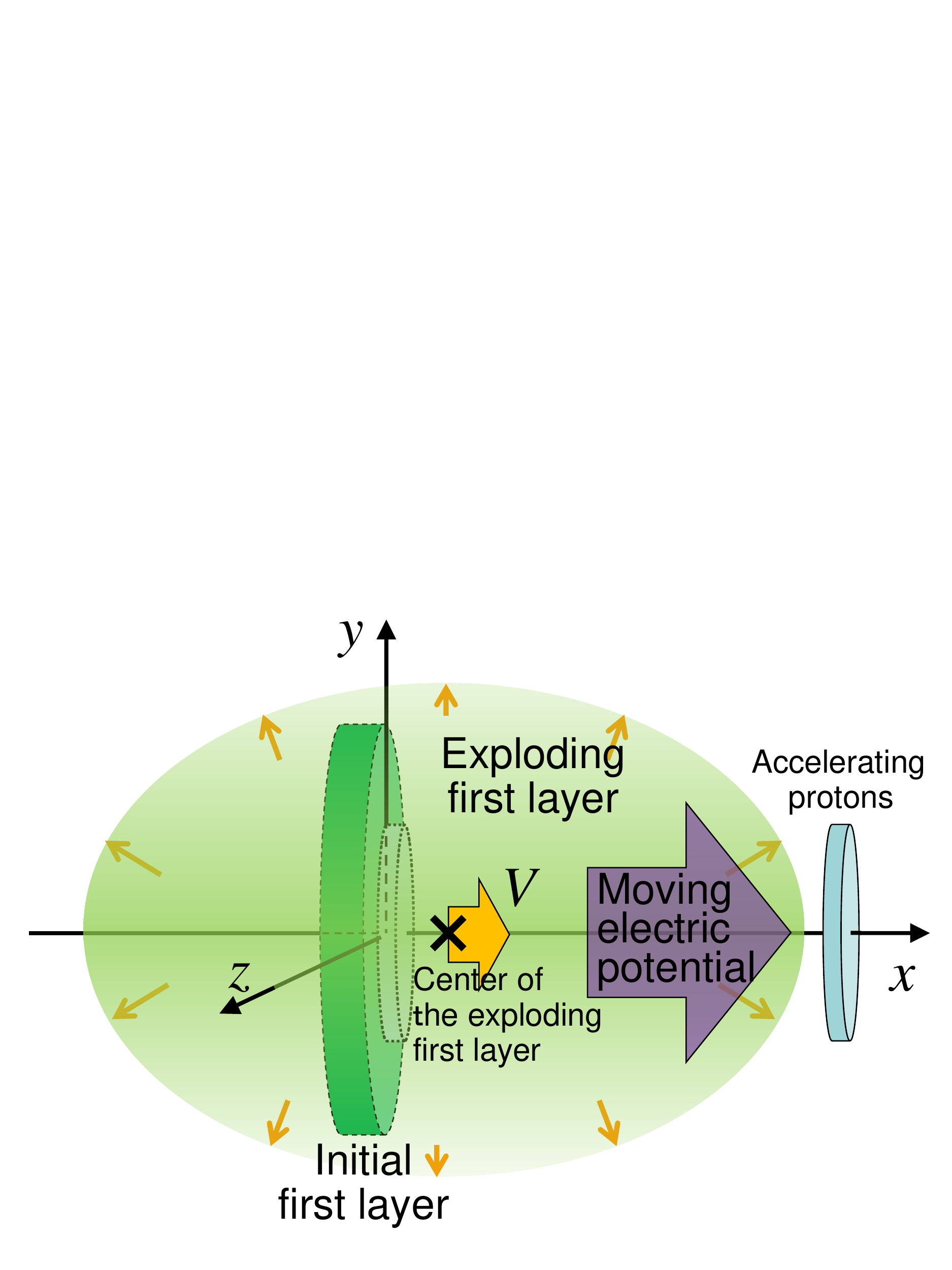}
\caption{
The first layer using ``light'' materials produces
a strongly inhomogeneous expansion due to the Coulomb explosion (light pattern).
The expanding first layer moves at average velocity $V$
in the direction of laser propagation by RPDA.
The electric potential moves in the $x$ direction as a result of these effects.
}
\label{fig:fig02}
\end{figure}

One way this situation can be created by the use of a Coulomb explosion
of the first layer.
Figure \ref{fig:fig02} shows that
the first layer disk undergoes a strongly inhomogeneous expansion
owing to the Coulomb explosion.
This expansion raises the moving electric potential for
the accelerating protons.
In other words,
many ions in the first layer are distributed close to the
accelerating protons keeping a comparatively high density
and move in the proton direction.
The acceleration rate is higher when
the expansion velocity of the first layer ions is higher.
This means that
the strong Coulomb expansion operates effectively for proton acceleration.
The Coulomb explosion level is determined by the ``mass''
of the ions composing the first layer.
Equation (\ref{emv}) shows that ``light'' ions have a high expansion velocity.
That is, ``light'' ions undergo a stronger a Coulomb explosion
and should be generating higher energy protons.

Another way to induce movement of the first layer
is by radiation-pressure-dominant acceleration (RPDA).
Figure \ref{fig:fig02} shows that the first layer,
which expands by a Coulomb explosion (with an ellipsoidal light pattern),
is moving with velocity $V$ in the laser propagation direction
(proton direction) by RPDA.
This movement leads the moving electric potential.
Higher $V$ values generate higher energy protons,
since the protons experience the accelerating electric field over a longer time
by following the electric potential.
A portion of the energy and momentum transferred from the laser pulse to the electrons
is imparted to the ions via a charge separation field.
That is, the ions get accelerated by this field,
and the ``light'' ions have higher velocity (Eq. (\ref{emv})).
Thus the ``light'' ions experience a higher first layer velocity
and should be generating higher energy protons.

One can obtain higher energy protons by using
a ``light'' material in the first layer, as
 is corroborated by the simulations described below.
The simulations were performed with a 3D massively parallel
electromagnetic code, based on the PIC method. \cite{CBL}

\section{SIMULATION OF FIRST LAYER MATERIALS} \label{sim-a}
In this section, I study the dependence of the proton energy
on the first layer materials of the double-layer target
by using simulations.

\subsection{Simulation parameters}

Here, I show the parameters used in the simulations.
The spatial coordinates are normalized by the laser wavelength
$\lambda=0.8$ $\mu$m
and time is measured in terms of the laser period, $2\pi/\omega$.

I use an idealized model, in which a Gaussian linearly polarized laser pulse is
incident on a double-layer target represented by a collisionless plasma.

The laser pulse with dimensionless amplitude
$a=q_eE_{0}/m_{e}\omega c=50$,
which corresponds to a laser peak intensity of $5\times 10^{21}$ W/cm$^{2}$,
is $10\lambda $ long in the propagation direction, $27$ fs in duration,
and focused to a spot with size $4\lambda $ (FWHM),
which corresponds to a laser peak power of $620$ TW and a laser energy of $18$ J.
Here, the laser peak power is calculated by using
$\int_{-\infty}^{\infty} \int_{-\infty}^{\infty} I(y,z)dydz$ and
the laser energy is calculated by using
$\int_{-\infty}^{\infty}\int_{-\infty}^{\infty}\int_{-\infty}^{\infty}I(y,z,t)dydzdt$,
where $I$ is the laser intensity.
The laser pulse is normally incident on the target.
The electric field is oriented in the $y$ direction.
I use this laser pulse in all simulations in this paper.

Both layers of the double-layer target are shaped as disks.
The first layer has a diameter of $8\lambda $ and a thickness of
$0.5\lambda $.
The second, hydrogen,
layer is narrower and thinner; its diameter is $4\lambda $
and its thickness is $0.03\lambda $.
The electron density inside the first layer is $n_{e}=3\times 10^{22}$ cm$^{-3}$
and inside the hydrogen layer it is $n_{e}=9\times 10^{20}$ cm$^{-3}$.
The total number of quasiparticles is $8\times 10^{7}$.
The number of grid cells is equal to $3300\times1024\times 1024$
along the $X$, $Y$, and $Z$ axes, respectively.
Correspondingly, the simulation box size is
$120\lambda \times 36.5\lambda \times 36.5\lambda$.
The boundary conditions for the particles and for the fields are
periodic in the transverse ($Y$,$Z$) directions and absorbing at the
boundaries of the computation box along the $X$ axis.
 $xyz$ coordinates are used in the text and figures;
the origin of the coordinate system is located at the center of the rear surface
of the initial first layer, and the directions of the $x$, $y$, and $z$ axes are the same as
those of the $X$, $Y$, and $Z$ axes, respectively.
That is, the $x$ axis denotes the direction perpendicular to the target surface
and the $y$ and $z$ axes lie in the plane of the target surface.

Although the first layer material can be varied,
the number of ions is the same in all cases and the ionization state of
each ion is assumed to be $Z_{i}=+6$, \cite{HSB,HKM,JHK}
since the laser parameters and the target geometry are fixed.

\subsection{Proton energy as a function of first layer materials}

\begin{figure}[tbp]
\includegraphics[clip,width=10.0cm,bb=0 26 479 690]{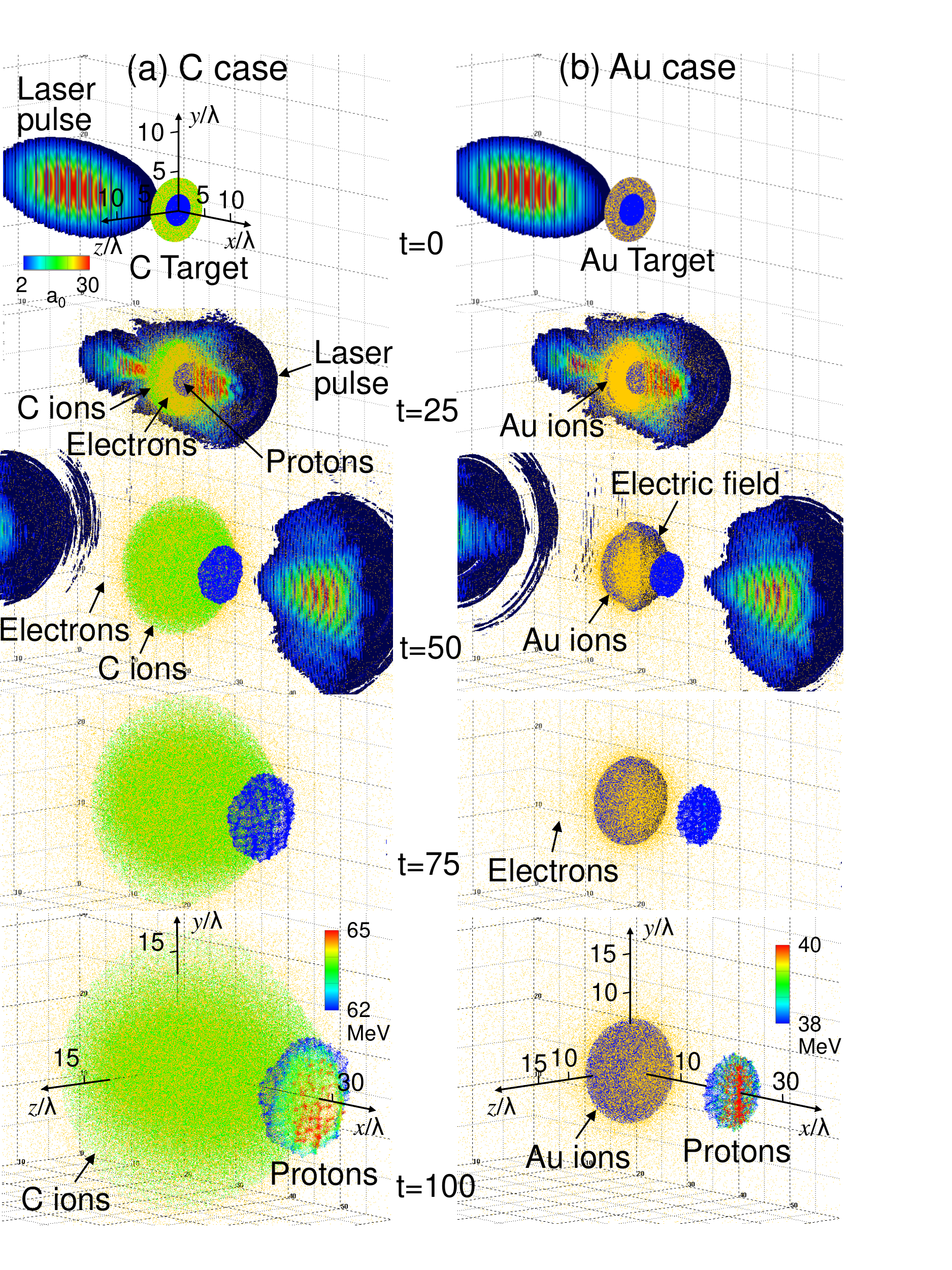}
\caption{
Particle distribution and electric field magnitude (isosurface for value $a=2$)
for carbon (a) and gold (b).
Half of the electric field box has been removed to reveal the internal
structure.
Shown are the initial shape of the target and the laser pulse ($t=0$),
the interaction of the target and laser pulse ($t=25,50\times 2\pi/\omega$),
and the first layer shape and the accelerated protons (color scale)
($t=75,100\times 2\pi/\omega$).
}
\label{fig:fig03}
\end{figure}

\begin{figure}[tbp]
\includegraphics[clip,width=8.0cm,bb=0 0 530 708]{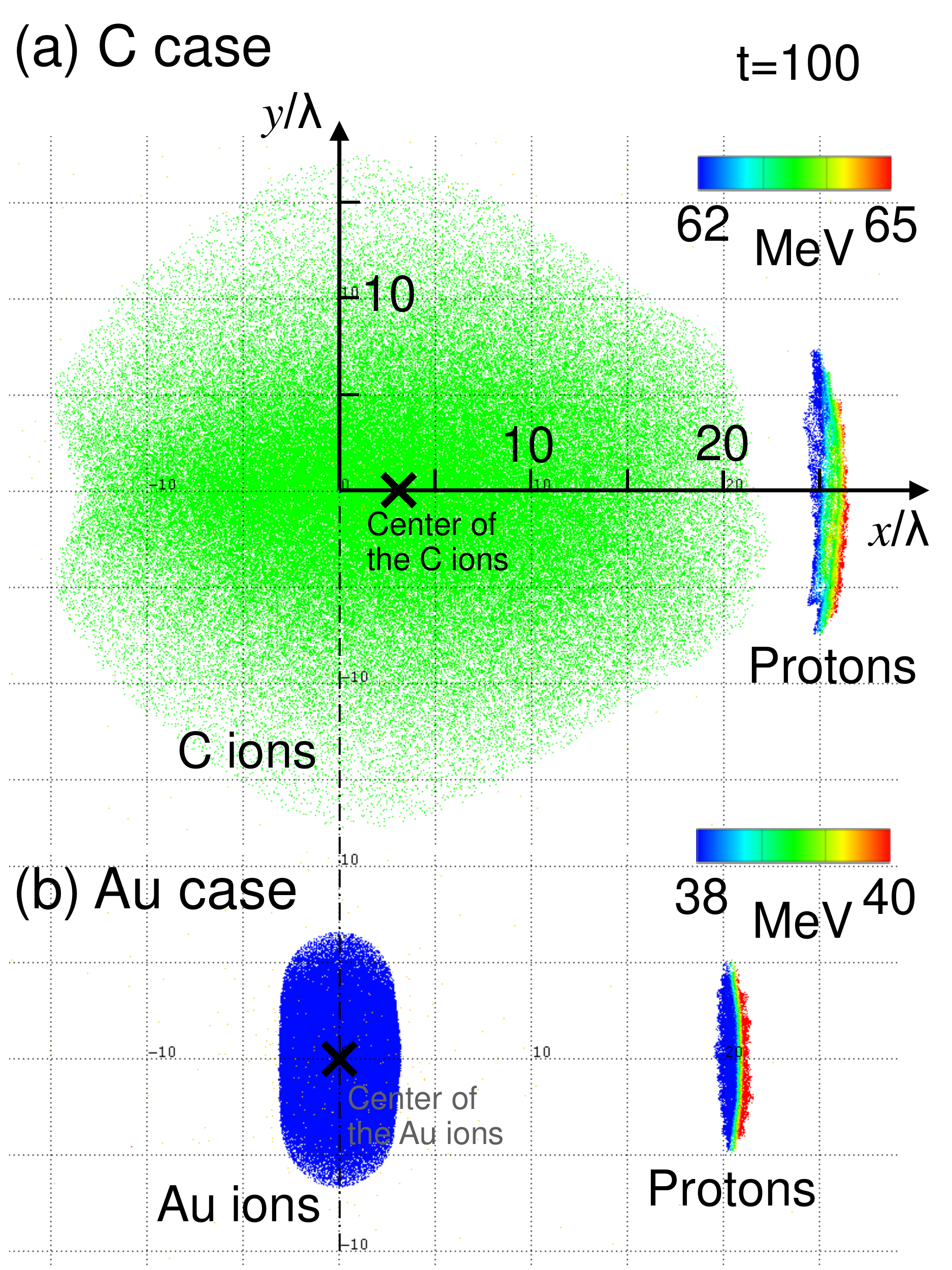}
\caption{
Distribution of the ions of the first layer and protons (color scale)
in the carbon (a) and gold (b) cases at $t=100\times 2\pi/\omega$;
a two-dimensional projection is shown looking along the $z$ axis.
In the carbon case,
the distribution area of the carbon ion cloud is very wide
and it moves in the $x$ direction.
}
\label{fig:fig04}
\end{figure}

\begin{figure}[tbp]
\includegraphics[clip,width=8.0cm,bb=20 0 527 438]{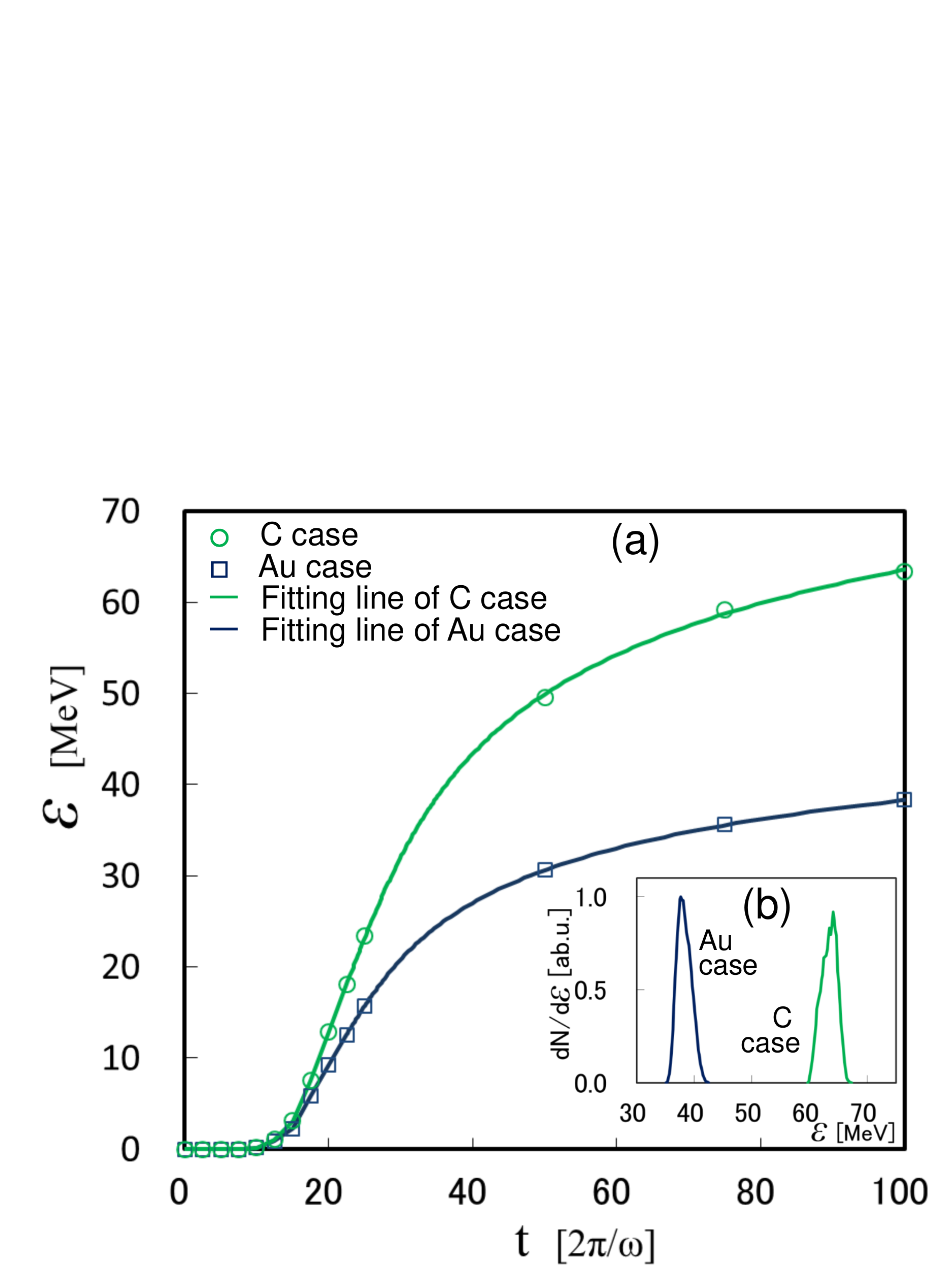}
\caption{
(a) Average proton energy $\mathcal{E}$ versus time,
as obtained in the simulation shown in Figs. \ref{fig:fig03}.
The proton energy of the carbon case is higher than in the gold case
at all times.
(b) Proton energy spectrum obtained in the simulation
at $t=100\times 2\pi/\omega$,
for the cases of gold and carbon,
normalized by the maximum in the former case.
}
\label{fig:fig05}
\end{figure}

\begin{figure}[tbp]
\includegraphics[clip,width=8.0cm,bb=0 0 520 444]{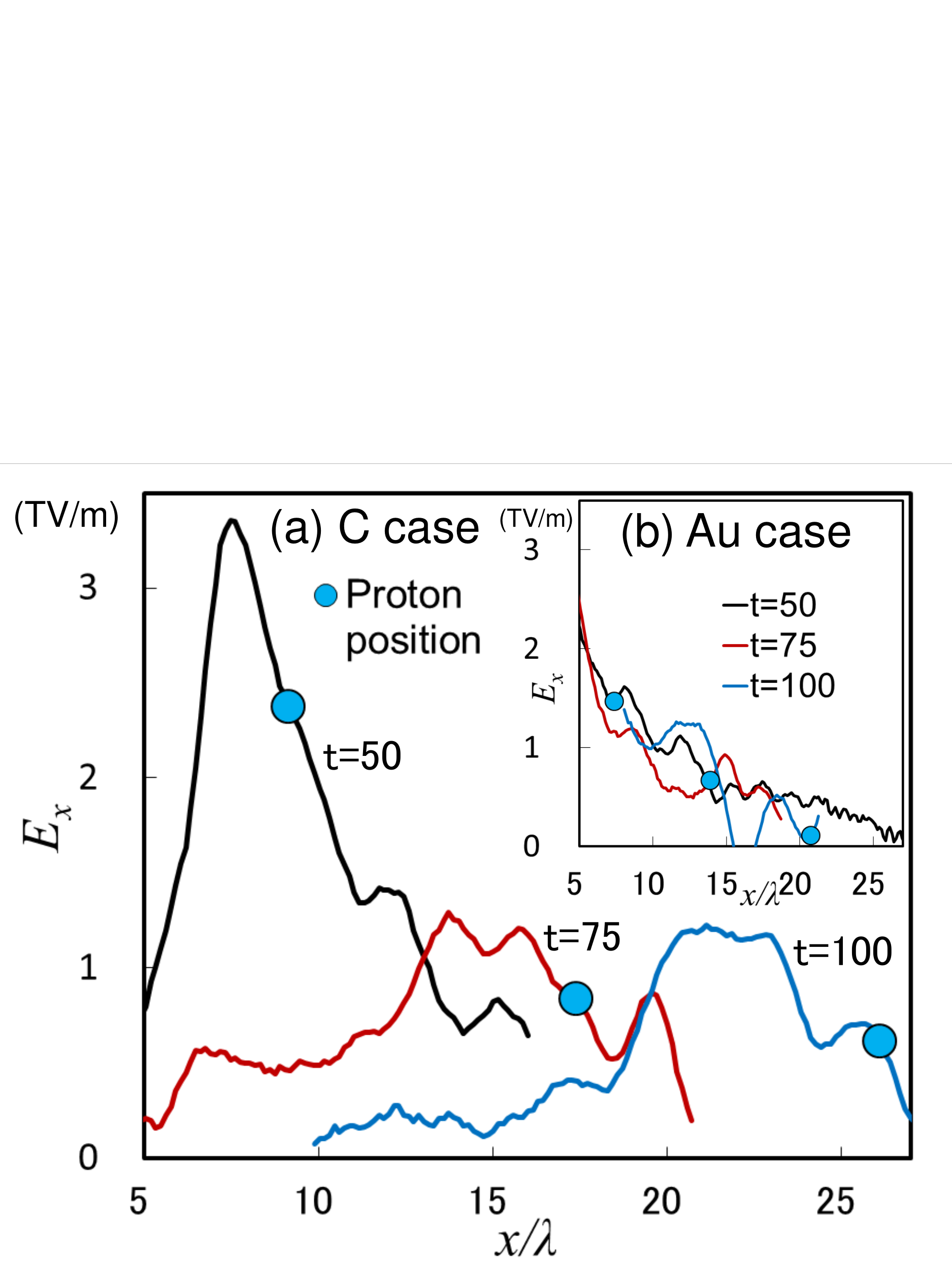}
\caption{
The $x$ component of the electric field, $E_x(x)$ (solid lines)
and the proton position (circles) at each time.
In the gold case, in the inset, the electric field simply decreases
with distance from the target surface.
In the carbon case, it does not decrease much and
has a peak that moves to the right side.
}
\label{fig:fig07}
\end{figure}

\begin{figure}[tbp]
\includegraphics[clip,width=13.0cm,bb=0 0 540 179]{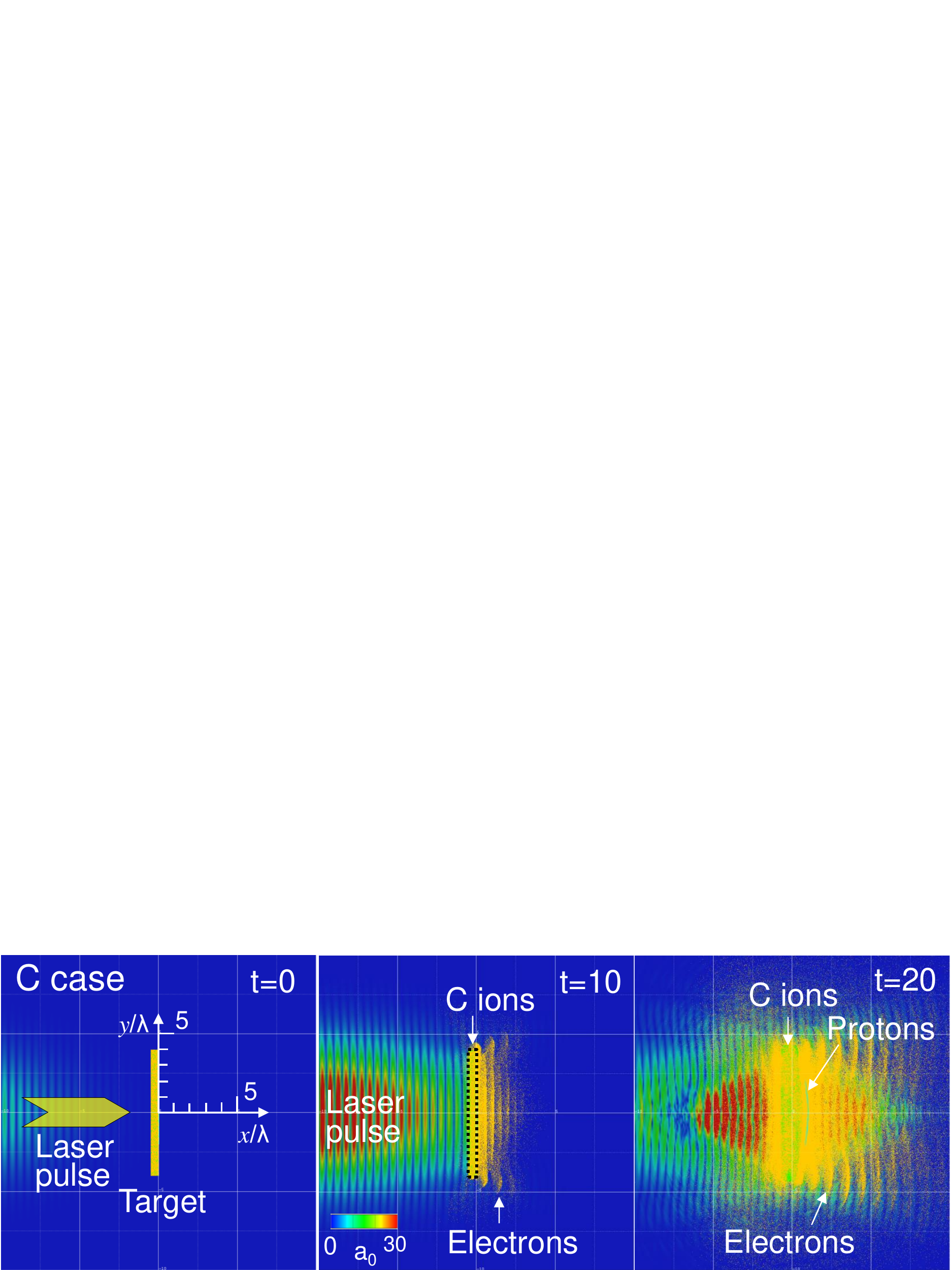}
\caption{
Spatial distribution of particles and the electric field magnitude
at early simulation times for the carbon case.
A two-dimensional projection is shown looking along the $z$ axis.
The solid bar is a projection of the first layer ($t=0$).
The gas pattern denotes electrons
that are pushed out from the target by the laser pulse
($t=10, 20\times 2\pi/\omega$) and
distributed to the rear area (propagation direction) of the target.
This produces the movement of the first layer ions.
}
\label{fig:fig08}
\end{figure}

\begin{figure}[tbp]
\includegraphics[clip,width=8.0cm,bb=7 2 528 450]{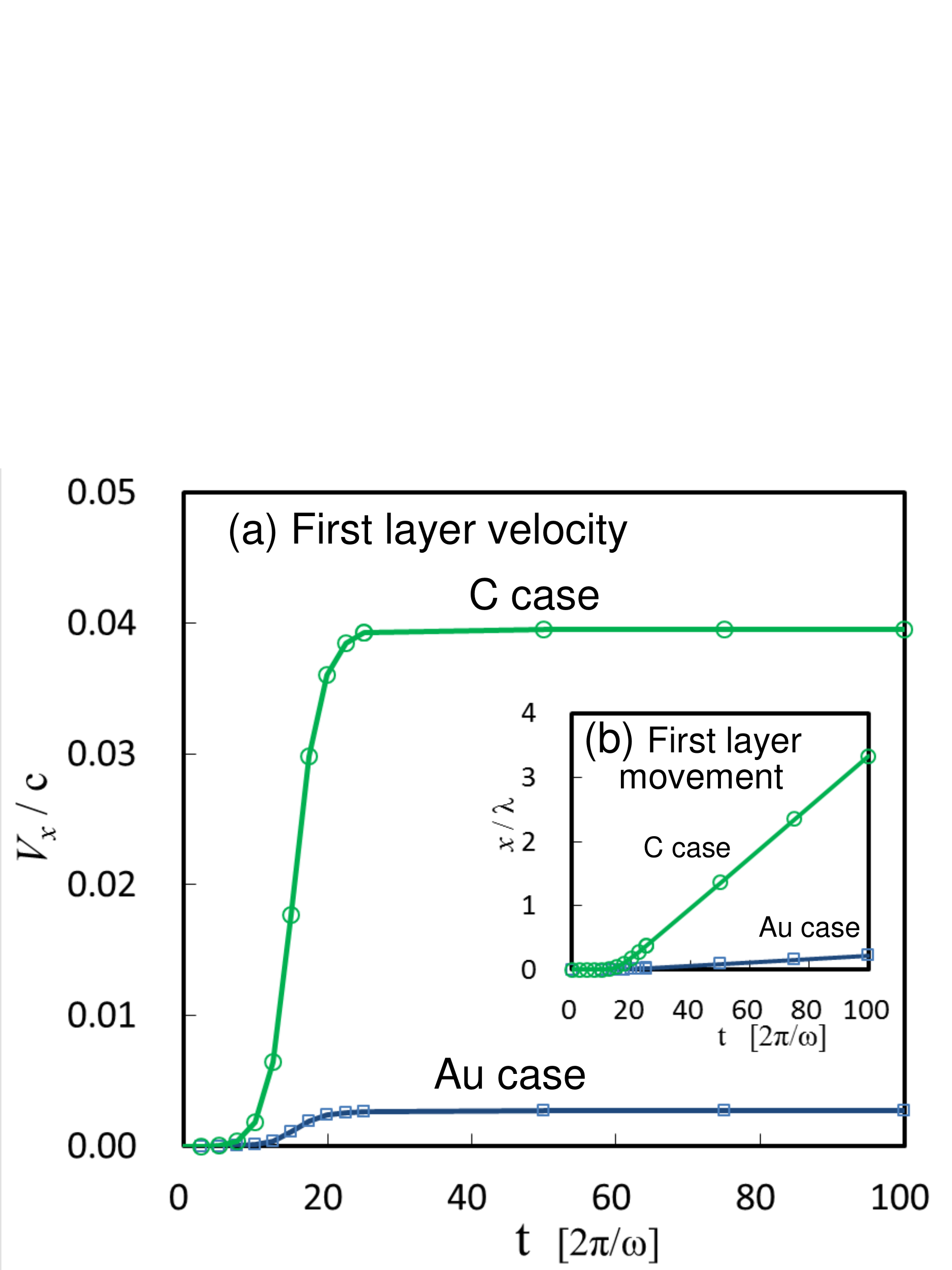}
\caption{
(a) Velocity of the first layer of the target in the $x$ direction
 normalized by the speed of light, $V_x/c$, as a function of time.
The velocity of the carbon target is much higher than that for the gold case.
(b) Movements of the first layer in the $x$ direction
 normalized by the wavelength, $x/\lambda$, as a function of time.
}
\label{fig:fig09}
\end{figure}

To examine the dependence of the proton energy, $\mathcal{E}$,
on the first layer material,
I performed simulations using different materials
at normal incidence to the laser pulse.
First, I show two simulation results, one in which the first layer consists of carbon (a ``light'' material)
and the other is in which it consists of gold (a ``heavy'' material).
The results are shown in Figs. \ref{fig:fig03}--\ref{fig:fig09}.

Figure \ref{fig:fig03} shows
the particle distribution and electric field magnitude of both cases
at each time.
We see that
the carbon ions are distributed over a much wider area by its Coulomb explosion
than are the gold ions.
However, the deformation of the laser pulse is similar
in both cases at all times.
The Coulomb explosion process of the first layer is much slower
than the laser pulse progress.
The first layer is almost undeformed at $t=25\times 2\pi/\omega$
when the laser pulse is just around the target and
it has the strong interactions with the target.
A big deformation of the first layer appears at $t>25\times 2\pi/\omega$
when the laser pulse passes through or reflects from the target.
Therefore, the explosion of the first layer is an almost simple
Coulomb explosion without other effects.
The proton energy obtained in the carbon case is much higher than that in the gold case.
The average proton energy at $t=100\times 2\pi/\omega$,
$\mathcal{E}_\mathrm{ave}$,
is $63$ MeV for the carbon case and $38$ MeV for the gold case
(a factor of 1.7 higher).

The reason why higher energy protons are obtained in the carbon case is that
the first layer deformation effectively contributes to the proton acceleration.
Let us examine the deformation of the first layer.
Figure \ref{fig:fig04} shows a cross section of the ion density distribution
near the $(x,y,z=0)$ plane at $t=100\times 2\pi/\omega$, as seen by
 looking along the $z$ axis.
In the carbon case, the strong Coulomb explosion distributes
the carbon ions ellipsoidally,
and the distribution area is much wider than in the gold case.
The expansion of the cloud of carbon ions is strongly inhomogeneous and
it appears to be substantially elongated in the longitudinal direction.
The surface of the carbon ion cloud is close to the acceleration protons,
which means that the protons keep getting pushed by a comparatively strong force.
Moreover, the center point of the ion cloud is moving in the $x$ direction,
at a coordinate of $3.1\lambda$,
so that the electrostatic potential is moving in the direction of the protons.
In contrast, in the gold case,
the ion distribution is very compact and
the distance between the ion cloud surface and the protons is much greater
than in the carbon case.
In addition, it almost does not move,
with the coordinate of the center point of gold ions being $0.0\lambda$.
Moreover,
the proton positions in the carbon case have travelled further along the
$x$ axis than in the gold case,
since the protons in the carbon case have higher energy
(higher velocity) than in the gold case.

Here, let us consider the variation of the proton energy in time for both cases.
Figure \ref{fig:fig05}(a) shows the average proton energy versus time.
The proton energy in the carbon case is always higher than in the gold case
and difference in energy grows in time.
The protons are accelerated relatively quickly until a time
of about $t=50\times2\pi/\omega$ in both cases.
In the gold case, the acceleration almost saturates at $t>50\times2\pi/\omega$,
although in the carbon case it is still increasing compared with the gold case.
This is because the electric field seen by protons
in the carbon case is greater and continues longer than in the gold case,
owing to the expansion and movement of first layer (see Fig. \ref{fig:fig04}).
Correspondingly, their energy gain is also greater.
In the carbon case the proton energy is 1.7 times higher than in the gold case
and the energy spread is almost the same in the both cases,
as seen in Fig. \ref{fig:fig05}(b).
The energy spread, $\Delta\mathcal{E}/\mathcal{E}_\mathrm{ave}$, is 6\%\
and 8$\%$ for the cases of carbon and gold, respectively.
Higher energy protons can be obtained in the carbon first layer,
as a result of the stronger electric field and movement of the first layer.
I consider the differences in the electric field and
the first layer movement in the two cases.

First, I show the differences in the electric field between the two cases.
Figure \ref{fig:fig07} shows the $x$ component of the electric field along
the $x$ coordinate (solid lines), $E_x(x)$,
and the proton position (circular dots) at each time.
In the gold case (see Fig. \ref{fig:fig07}(b)),
the electric field simply decreases with the distance from the
surface of the initial target at all times.
Therefore, the electric field that protons experience (see the circular dots)
simply decreases too.
In contrast, in the carbon case (Fig. \ref{fig:fig07}(a)),
the electric field peaks at points near the protons for each time.
The electric field moves in the direction of the accelerating protons
keeping the same shape (see Fig. \ref{fig:fig07}(a): $t=75$
and $t=100\times2\pi/\omega$).
The electric fields at the proton positions slowly decrease,
and those are higher than in the gold case at all times.
At the proton position at $t=50\times2\pi/\omega$,
the value of $E_x$ is $2.4$ and $1.4$ TV/m
for the cases of carbon and gold ions, respectively.

Next, I consider the movement of the first layer.
Figure \ref{fig:fig08} shows the laser pulse and the target at an early time,
$t<20\times2\pi/\omega$, for the carbon case, as
seen by looking along the $z$ axis.
The laser pulse moves from the left-hand size to the right-hand side,
and the color shows the electric field magnitude.
We can see that many electrons (gas pattern) are pushed out
from the target by the ponderomotive force from the laser pulse.
Those pushed-out electrons are distributed to
the rear area (propagation direction) of the target
and they move forward in the laser propagation direction,
although the carbon ions stay at their initial position.
This charge separation produces the strong electric field.
Then, the carbon ions are moved in the laser propagation direction
by this electric field (which is RPDA).
Figure \ref{fig:fig09} shows
the first layer velocity, $V_x$, normalized by the speed of light,
and the first layer position, normalized by the wavelength,
for the $x$ direction as a function of time.
These are averaged values for all ions of the first layer.
The first layer velocity rises rapidly at the initial time,
$t \sim 20\times2\pi/\omega,$
when the laser pulse is still around the target (see Fig. \ref{fig:fig08})
and the velocity is constant at time $t>25\times2\pi/\omega,$
 after the laser pulse passes through or reflects off the target.
The increase in the fist layer velocity stops at
$t \approx 20\times 2\pi/\omega$, since by this time
there is no clearly one-sided distribution of the electrons like that at
time $t \approx 10\times2\pi/\omega$ (see Fig. \ref{fig:fig08}).
The first layer velocity in the carbon case is $14$ times that in
the gold case at $t>25\times 2\pi/\omega$.
This means strong RPDA occurs in the carbon case.
The movement in the carbon case is much greater than in the gold case too,
and the difference in distance between the carbon case and the gold case grows
with time.
This indicates that the moving electric potential operates efficiently in the carbon case.
Incidentally, the velocity for the $y$ direction, $V_y$, of the first layer
is relatively very small.
It is about $1/1000$ of $V_x$ in both the carbon and gold cases.
Movement for the $y$ direction is very small too,
because they are normal to the incidence of the laser pulse.

\begin{figure}[tbp]
\includegraphics[clip,width=7.0cm,bb=25 3 518 446]{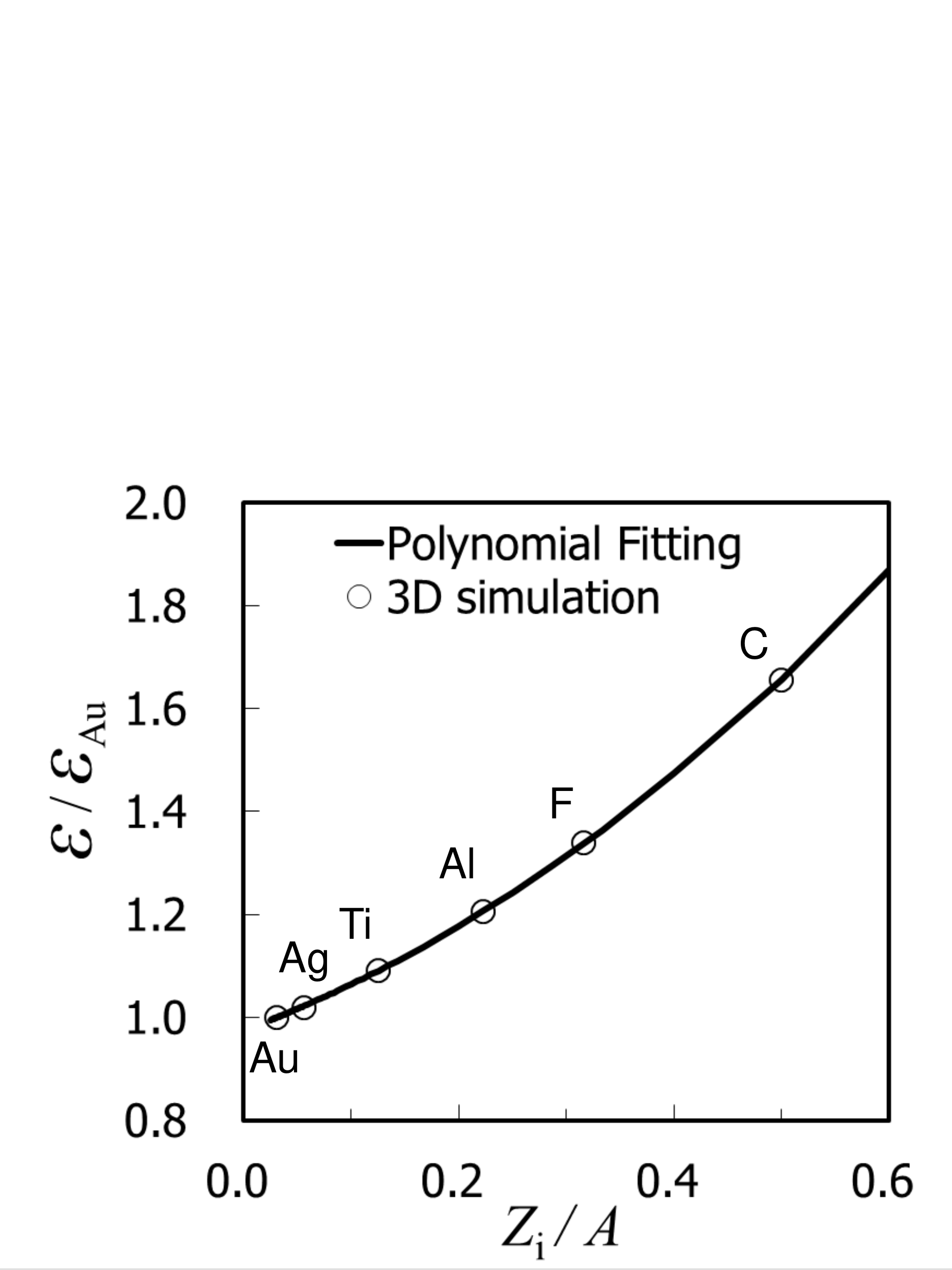}
\caption{
Average proton energy, normalized by energy in the gold case,
shown with respect to the ratio between the charge state, $Z_{\rm i}$,
and the atomic number, $A$, of ions making up the first layer.
}
\label{fig:fig10}
\end{figure}

In above considerations,
I showed that the higher energy protons can be obtained
for ``lighter'' material in the first layer
by comparing carbon and gold.
Here, I investigate the effect using additional materials.
Figure \ref{fig:fig10} shows the average proton energy for different
materials comprising the first layer.
The proton energy is normalized by the energy in the case of gold ions.
We see that the average proton energy almost linearly depends
on the ratio of $Z_{\rm i}/A$,
where $Z_{\rm i}$ is the charge state and $A$ is the atomic mass number
of the ion.
Higher energy protons can be obtained by using a larger ratio of $Z_{\rm i}/A$,
i.e., ``light,'' material for the first layer in the double-layer target.

In experiments on laser-driven ion acceleration,
 a CH polymer target exhibited higher
energy protons \cite{SNAV}
than a metallic target. \cite{CLAR}

\section{COMPONENTS OF THE PROTON ENERGY} \label{sim-b}

In the previous section, I showed that higher energy protons can be obtained
by using ``light'' material in the first layer, because a strong Coulomb explosion and a greater first layer movement
toward the accelerating protons occur in such materials.
These effects augment the acceleration by the electric field
of the charged first layer disk.
In this section, I show the amount of each effect of the acceleration
on the total proton energy in the carbon case.

\begin{figure}[tbp]
\includegraphics[clip,width=10.0cm,bb=0 25 474 677]{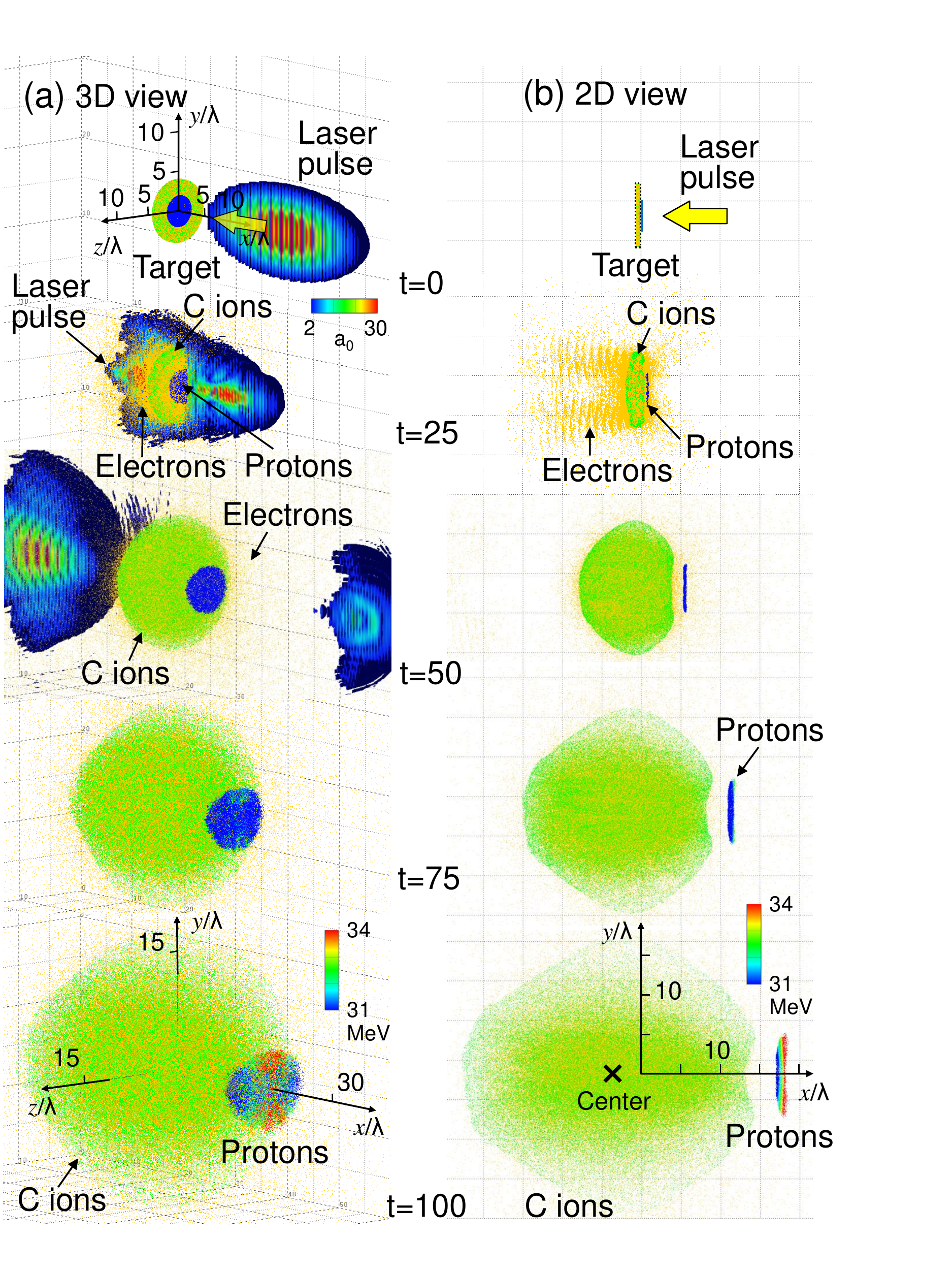}
\caption{
(a) 3D view and (b) the projection onto the $(x,y)$ plane
for a laser pulse irradiating onto the other side, the
proton layer side, of a double-layer target.
Shown are the particle distribution and electric field magnitude
(isosurface for value $a=2$); half of the electric field box has been removed.
The protons are accelerated in the $+x$ direction,
and the C ions move in the $-x$ direction by RPDA.
}
\label{fig:fig11}
\end{figure}

To examine the effect of the first layer velocity, $V$,
I performed simulations with a laser pulse by reversing the irradiation
direction of the laser (reverse irradiating)
in the previous carbon case.
Thus the hydrogen layer was put on the front side of the target
(see Fig. \ref{fig:fig11}).
\begin{figure}[tbp]
\includegraphics[clip,width=7.0cm,bb=0 0 538 370]{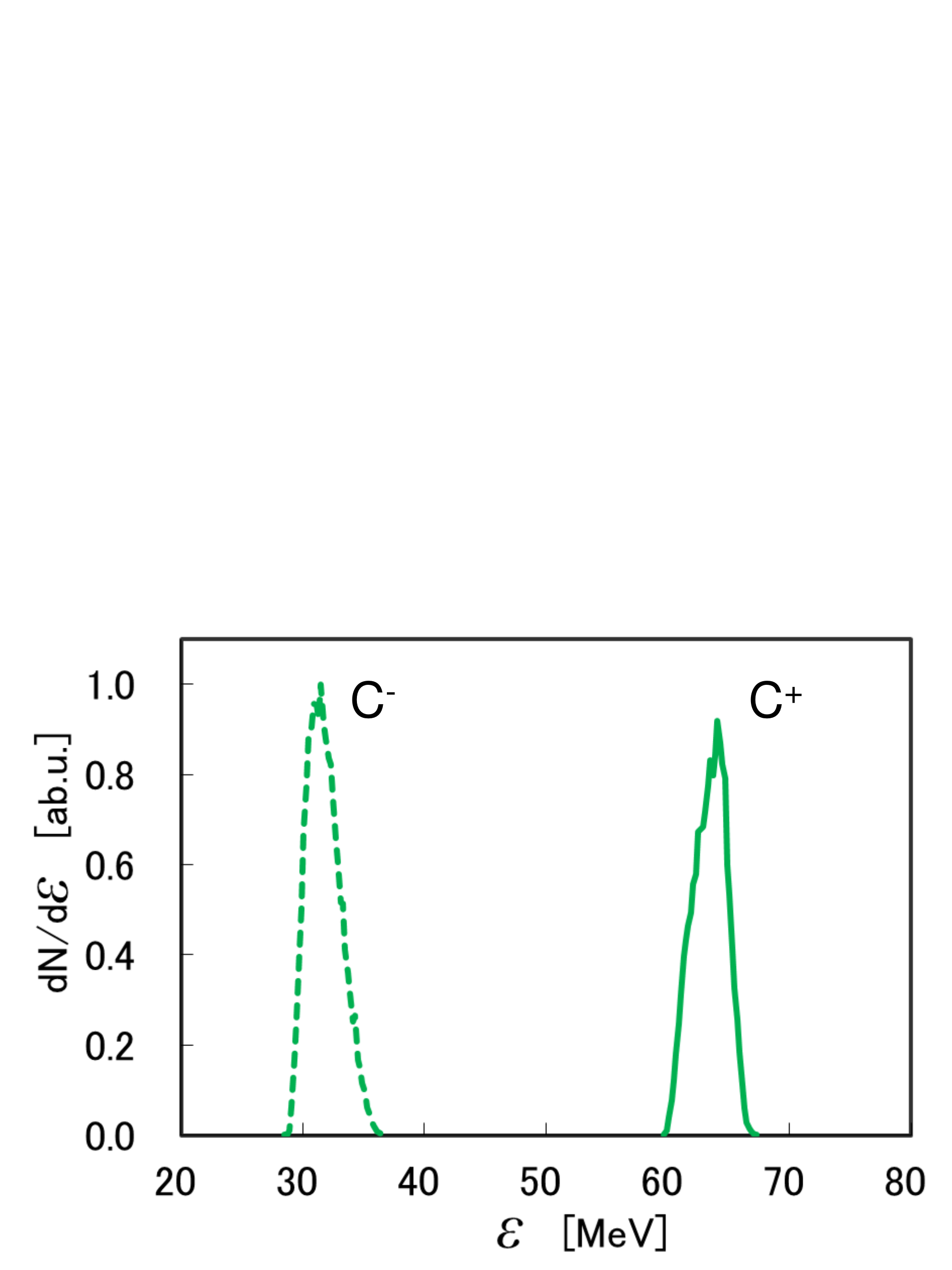}
\caption{
Proton energy spectrum obtained in the simulation
at $t=100\times 2\pi/\omega$, for the reverse irradiation case, $C^-$,
and the positive irradiation (previous carbon) case, $C^+$,
normalized by the maximum in the former case.
}
\label{fig:fig12}
\end{figure}
The results are shown in Figs. \ref{fig:fig11} and \ref{fig:fig12}.
The protons are accelerating in the $+x$ direction
even when the laser pulse irradiates the target in the reversed way.
The average proton energy in this case, $\mathcal{E}^{-}$, is $32$ MeV.
Since the proton layer is very thin and small,
it has less effect on the first layer velocity.
Therefore, the first layer velocity, $V^-$, has the same absolute value as in
positive irradiation, the case of the previous section,
but the sign is opposite, $V^-(t)=-V(t)$,
where $V(t)$ is the first layer velocity in time of the positive case.
Therefore, the first layer movement has a negative effect on the proton energy
in the reverse irradiating case.
Using this result,
we can estimate the amount of proton energy attributable to
the first layer velocity, $\mathcal{E}_{V}$.
The proton energy in the positive irradiating case, $\mathcal{E}^{+}$,
and in the reverse irradiation case, $\mathcal{E}^{-}$, are written as
\begin{equation}
 \mathcal{E}^{+}=\mathcal{E}_a+\mathcal{E}_V,
\label{evp}
\end{equation}
\begin{equation}
 \mathcal{E}^{-}=\mathcal{E}_a-\mathcal{E}_V,
\label{evm}
\end{equation}
where $\mathcal{E}_a$ is the proton energy
without the work done by the first layer velocity
(i.e., attributable to the acceleration by the electric field of the charged disk of
the first layer and the Coulomb explosion of the first layer ions).
Hence we obtain the work done by the first layer velocity,
$\mathcal{E}_{V}=(\mathcal{E}^{+}-\mathcal{E}^{-})/2$, which is $16$ MeV.
Therefore, the ratio of the effect of the first layer velocity, RPDA, is $25\%$
of the total proton energy in the carbon case.
In our simulations, the laser intensity and energy of
$I_0=5\times 10^{21}$W/cm$^{2}$ and $\mathcal{E}_{las}=18$ J, respectively,
are not enough for the RPDA regime in full scale, but this shows that
the RPDA regime has a strong effect even at this laser power level.
The proton energy without the work done by the first layer velocity,
$\mathcal{E}_a$, is $48$ MeV.

\begin{figure}[tbp]
\includegraphics[clip,width=10.0cm,bb=0 0 473 422]{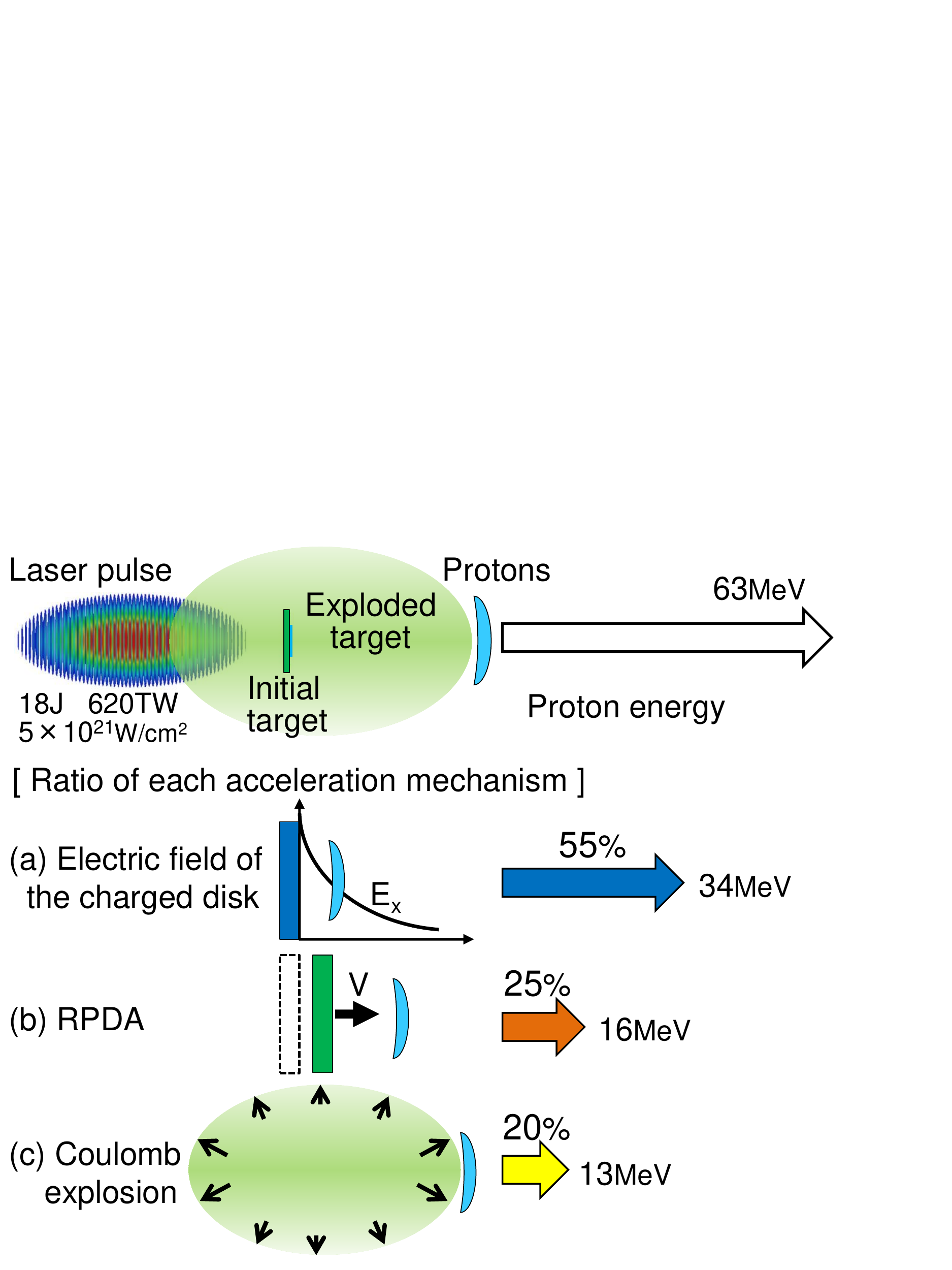}
\caption{
Proton energy of each acceleration mechanism in the carbon case.
The proton energy produced by the electric field of the nonmoving first layer
is almost half the total proton energy, and
RPDA and the Coulomb explosion each amount to almost a quarter of the total
proton energy.
}
\label{fig:fig13}
\end{figure}

Next,
I examine the work done by the Coulomb explosion of the first layer.
To do so,
I performed a simulation with a laser pulse reverse irradiating
in the gold case.
In this case, the average proton energy,
$\mathcal{E}^{'-}$, is $32$ MeV.
In the positive irradiation case, the gold case of the previous section,
the average proton energy, $\mathcal{E}^{'+}$, was $38$ MeV.
The difference between these values is very small,
because the first layer velocity, $V$, is very small
(see Figs. \ref{fig:fig04} and \ref{fig:fig09}).
This means that the work done by the first layer velocity is negligibly small
in the gold case.
The Coulomb explosion effect is negligible, as shown in previous section, too.
Therefore,
we estimate that the work without the first layer velocity
and Coulomb explosion in the carbon case,
$\mathcal{E}_0 \approx (\mathcal{E}^{'+}+\mathcal{E}^{'-})/2$, is $35$ MeV.
This is almost the energy from acceleration by only the electric field
of the charged disk of the first layer.
The ratio of this effect in the total energy of the carbon case is $55\%$.
Then we obtain the work done by the Coulomb explosion as
$\mathcal{E}_C=\mathcal{E}_a-\mathcal{E}_0=13$ MeV.
The ratio of the Coulomb explosion effect, $\mathcal{E}_{C}/\mathcal{E}^{+}$,
is $20\%$ in the carbon case
(see Fig. \ref{fig:fig13}).

Another effect to consider is that of the protons being dragged by the electrons
that are pushed out of the target by the laser pulse.
I estimate this effect by using the gold case results.
The electrons are mainly pushed out in the laser propagation direction.
Therefore,
in the reverse irradiation case,
the dragging effect by electrons acts to reduce the proton energy.
The proton energy in the positive irradiating case, $\mathcal{E}^{'+}$,
and in the reverse irradiation case, $\mathcal{E}^{'-}$, are written as
$\mathcal{E}^{'+}=\mathcal{E}^{'}_a+\mathcal{E}^{'}_V+\mathcal{E}^{'}_d$ and
$\mathcal{E}^{'-}=\mathcal{E}^{'}_a-\mathcal{E}^{'}_V-\mathcal{E}^{'}_d$,
where $\mathcal{E}^{'}_V$ is the work done by the first layer velocity,
$\mathcal{E}^{'}_d$ is the work done by the dragging by the electrons,
and $\mathcal{E}^{'}_a$ is other work.
The effect of the first layer velocity plus the dragging by electrons,
$\mathcal{E}^{'}_V+\mathcal{E}{'}_d=(\mathcal{E}^{'+}-\mathcal{E}^{'-})/2$,
is $3$ MeV.
Since $\mathcal{E}^{'}_V>0$,
the effect of the dragging by the electrons is $\mathcal{E}^{'}_d<3$ MeV,
and the ratio is less than $5\%$ of the total energy of the carbon case.
Therefore,
the effect whereby the protons are dragged by the electrons is very small
in our simulations.
This is because
the electrons move very fast compared with the ions and protons,
the distance between electrons and protons quickly becomes very large
compared with that between the first layer ions and protons,
and, moreover, the electrons are distributed over a very wide area.
The electric force decreases with distance by second order.
Therefore, the force the electrons exert on the protons is very small
compared with the force exerted by the ions.

Next,
I consider the theory for the work done by the first layer velocity.
I assume that the proton velocity $v > V$ and $v^2 \ll c^2$.
The theoretical formula \cite{MBEKK} is written as
$\mathcal{E}_V=mV^2(1+\sqrt{2\mathcal{E}_0/mV^2+1})$,
where $m$ is the proton mass and $\mathcal{E}_0$ is the proton energy in the
case of a nonmoving first layer.
Denoting the proton velocity in the case of a nonmoving first layer by $v_0$
and assuming $v_0^2 \ll c^2$ we obtain
\begin{equation}
 \tilde{\mathcal{E}}_V=2\tilde{V}(\tilde{V}+\sqrt{1+\tilde{V}^2}),
\label{eve0}
\end{equation}
where
$\tilde{\mathcal{E}_V}$ is the normalized proton energy
of the work done by the first layer velocity,
$\mathcal{E}_V/\mathcal{E}_0$,
and $\tilde{V}$ is the normalized velocity of the first layer, $V/v_0$.
\begin{figure}[tbp]
\includegraphics[clip,width=8.0cm,bb=15 0 522 442]{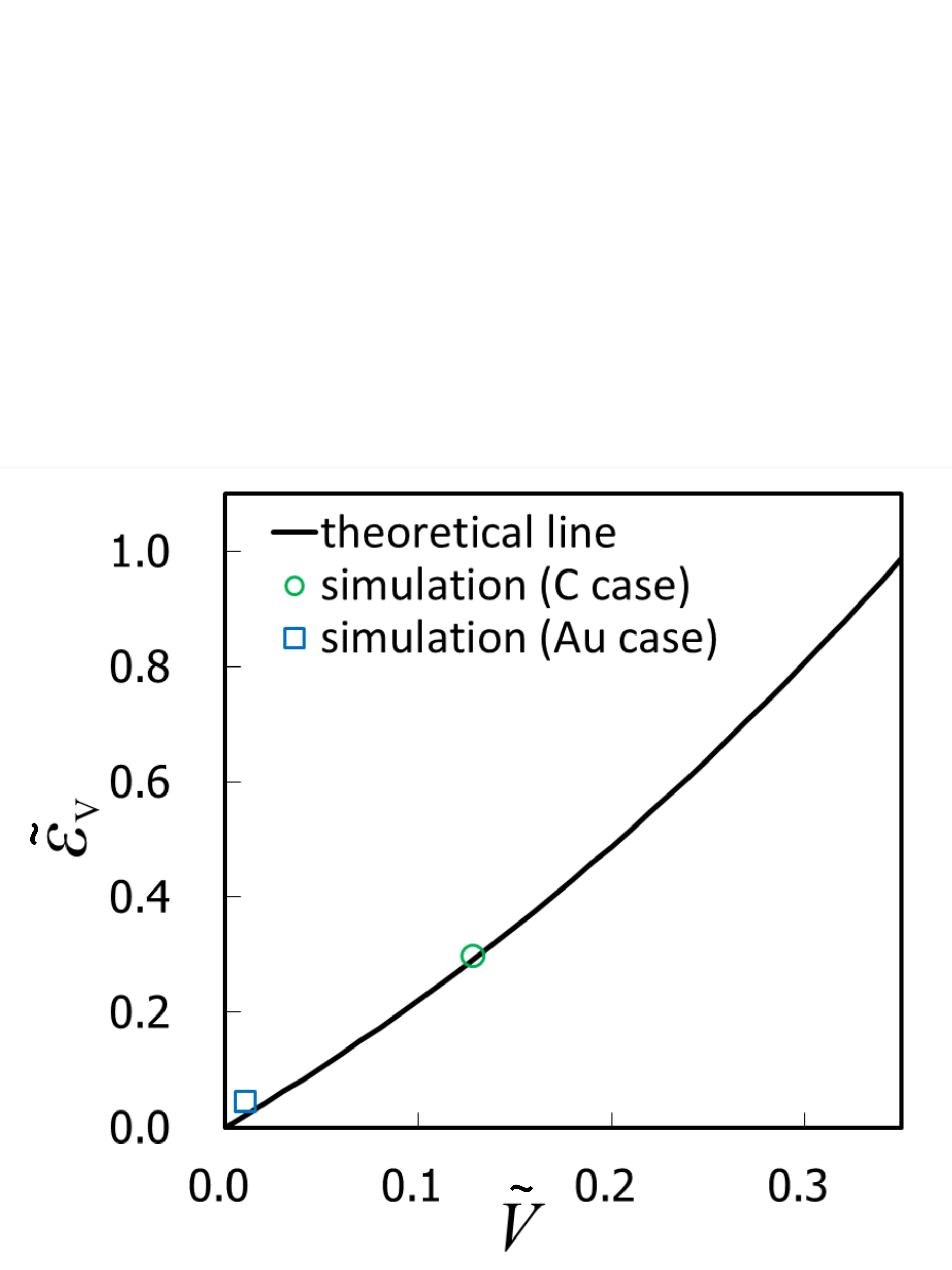}
\caption{
Proton energy produced by RPDA, $\tilde{\mathcal{E}}_V$,
as a function of first layer velocity, $\tilde{V}$.
The proton energy, $\mathcal{E}_V$,
is normalized by the proton energy of the nonmoving first layer case,
$\mathcal{E}_V/\mathcal{E}_0$,
and the first layer velocity, $V$, is normalized by
the proton velocity of the nonmoving first layer case, $V/v_0$.
The theoretical result (solid line) is given by equation (\ref{eve0}).
The simulation results for the gold and carbon cases are plotted with a
square and a circle, respectively.
}
\label{fig:fig14}
\end{figure}
In Fig. \ref{fig:fig14}, I present this theoretical
dependence of $\tilde{\mathcal{E}}_V$ on $\tilde{V}$,
by using formula (\ref{eve0}).
The proton energy by RPDA, $\mathcal{E}_V$, grows rapidly with increasing
 first layer velocity $V$.
The simulation results (circle and square dot) are plotted on the figure.
The theoretical calculations and the simulation results agree well.
Since the first layer velocity in the carbon case is not so high,
$V/v_0=0.13$,
Fig. \ref{fig:fig14} shows that
if we produce a high velocity in the first layer,
protons with a few times higher energy can be obtained.
In addition, assuming $\tilde{V}^2 \ll 1$ we obtain the simpler formula
$\tilde{\mathcal{E}}_V=2\tilde{V}$.

\section{Generating A 200-MeV proton beam}

I show the way to obtain a $200$-MeV proton beam,
using the same laser pulse as in the previous section
with the same normal incidence.
In the previous section,
I showed that higher energy protons can be obtained by using ``light'' material
in the first layer. The ``lightest'' material is hydrogen.
Therefore, we could get higher energy protons by using hydrogen
for the first layer.
I evaluate this contribution by simulation.
In this case,
even if we put the second layer of thin hydrogen on the hydrogen first layer,
the second layer has no meaning.
Therefore, I use a simple hydrogen disk, without a second layer.

The hydrogen disk target has the same shape and size as that of the
first layer of the double-layer target used previously.
The electron density inside the target is $n_{e}=9\times 10^{22}$ cm$^{-3}$.
Because the hydrogen cloud generated by the target is distributed over
a wider area than in the previous case,
I define a wider simulation box for $Y$ and $Z$ directions.
The number of grid cells is equal to $5000\times 3000\times 3000$ along the
$X$, $Y$, and $Z$ axes, respectively.
Correspondingly,
the simulation box size is $113\lambda \times 67.5\lambda \times 67.5\lambda$.
The total number of quasiparticles is $4\times 10^{8}$.
The other simulation parameters are the same as those used in previous sections.

\begin{figure}[tbp]
\includegraphics[clip,width=10.0cm,bb=0 0 520 695]{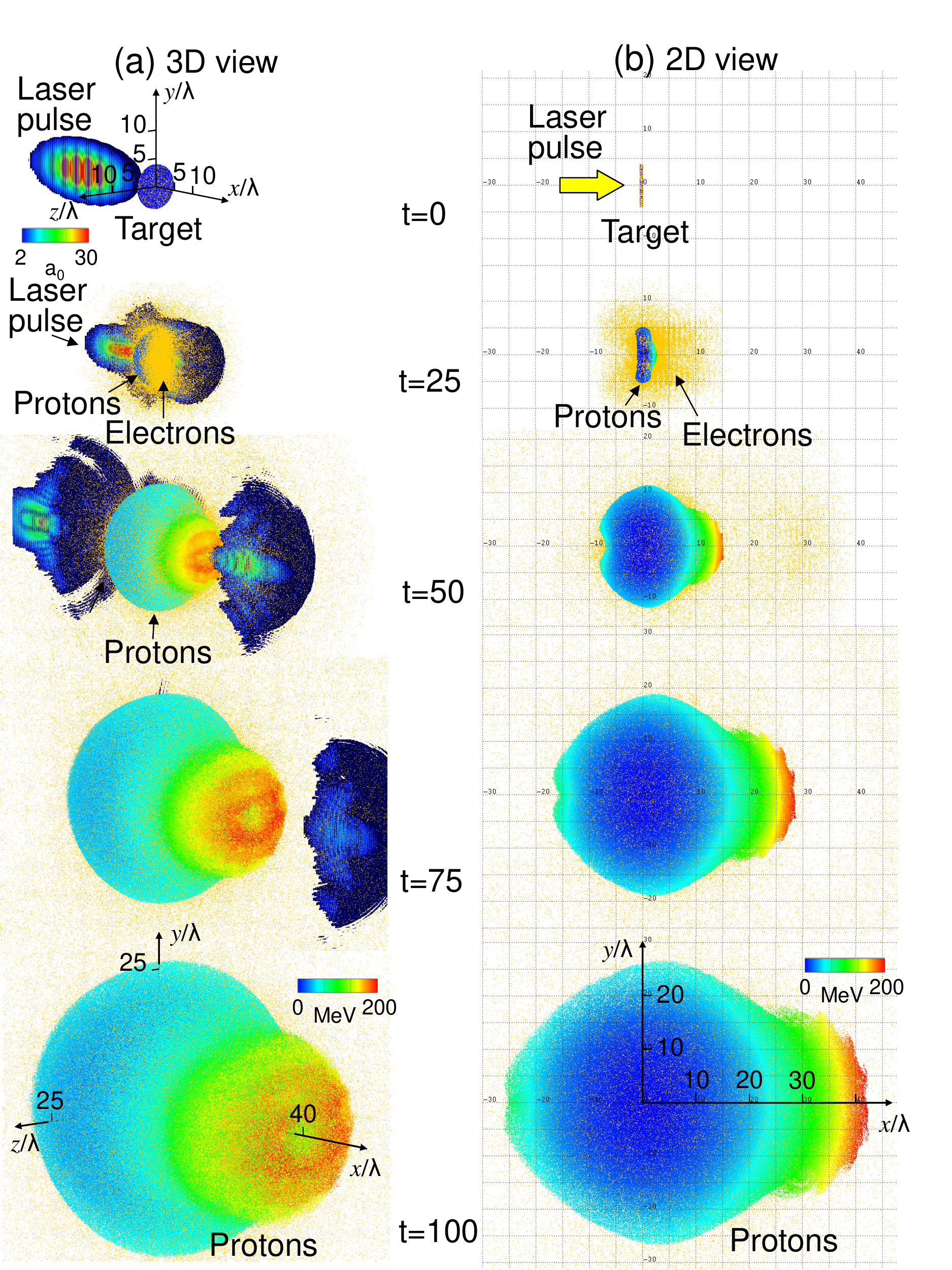}
\caption{
(a) Particle distribution and electric field magnitude
(isosurface for value $a=2$),
showing the initial shape of the target and the laser pulse ($t=0$) and
the interaction of the target and laser pulse ($t=25,50\times 2\pi/\omega$).
Half of the electric field box has been removed
to reveal the internal structure.
For protons, the color corresponds to energy.
(b) Two-dimensional projection of the particle distribution,
shown looking along the $z$ axis.
Half of the proton cloud has been removed
to reveal the internal structure.
}
\label{fig:fig-h1}
\end{figure}

Figure \ref{fig:fig-h1}(a) shows the particle distribution and
the electric field magnitude in time.
At $t=25\times 2\pi/\omega$, the laser pulse is just around the target and
it has the strong interactions with a target.
The target maintains its initial disk shape at this time.
After $t=25\times 2\pi/\omega$,
the laser pulse passes through or reflects off of the target,
and the proton cloud produced by Coulomb explosion is growing in time.
Figure \ref{fig:fig-h1}(b) shows a cross section of the ion cloud at each time.
Hydrogen ions (protons) are classified by color in terms of energy.
We see that the exploded hydrogen disk (hydrogen ions)
is distributed over a very wide area.
The expansion of the cloud of hydrogen ions is inhomogeneous and
the cloud is elongated in the longitudinal direction.
\begin{figure}[tbp]
\includegraphics[clip,width=8.0cm,bb=17 3 528 405]{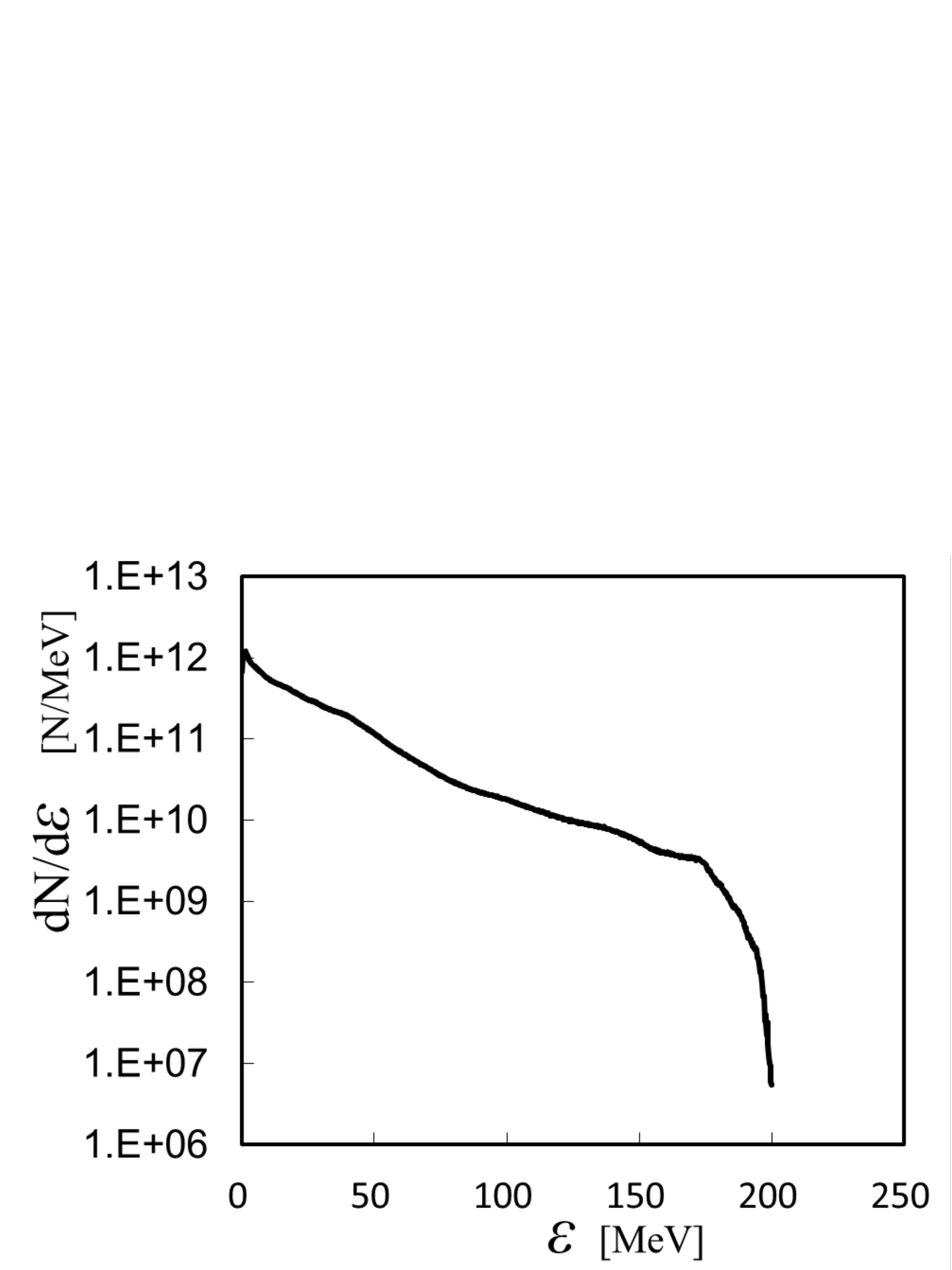}
\caption{
Proton energy spectrum obtained in the simulation
at $t=100\times 2\pi/\omega$.
}
\label{fig:fig-h2}
\end{figure}
Figure \ref{fig:fig-h2} shows the proton energy spectrum
at $t=100\times 2\pi/\omega$.
The maximum energy is $\mathcal{E}_\mathrm{max}=200$ MeV and
the average energy is $\mathcal{E}_\mathrm{ave}=25$ MeV.
The vertical axis is given in units of number of protons per $1$ MeV.
We can estimate the number of obtained protons by using the required
energy and energy width from Fig. \ref{fig:fig-h2}.
We see that we can obtain five times higher proton energy than in
the gold case,
even by using the same laser pulse,
by using an optimum target material.

\begin{figure}[tbp]
\includegraphics[clip,width=8.0cm,bb=7 2 526 451]{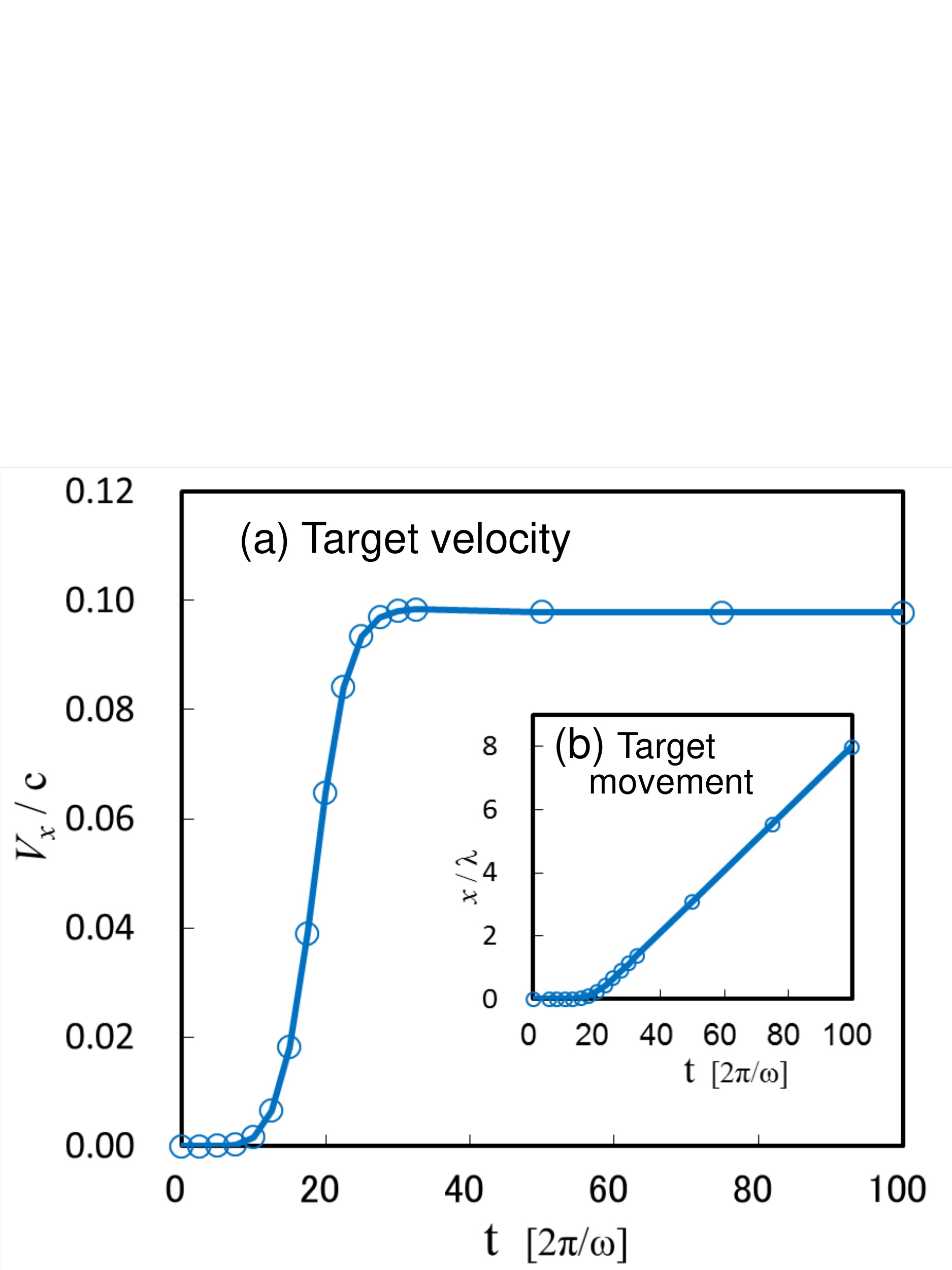}
\caption{
(a) Velocity of the hydrogen target in the $x$ direction,
 normalized by the speed of light, $V_x/c,$ as a function of time.
(b) Movement of the hydrogen target in the $x$ direction,
normalized by the wavelength, $x/\lambda$, as a function of time.
}
\label{fig:fig-h3}
\end{figure}

Figure \ref{fig:fig-h3} shows
the target average velocity, normalized by the speed of light,
and the target average position, normalized by the wavelength,
in the $x$ direction as a function of time.
The target velocity rises rapidly at the initial time,
$t \sim 20\times2\pi/\omega,$
when the laser pulse is still around the target
and the velocity is constant at time $t>25\times2\pi/\omega,$
after the laser pulse passes through or reflects off of the target.
This is similar to the carbon case (Fig. \ref{fig:fig09}),
although the target velocity of this case is $2.5$ times that in
the carbon case at time $t>25\times 2\pi/\omega$.
That means strong RPDA appears in this case,
because the target velocity is attributable to RPDA.
The movement of the target is much greater than in the carbon case too.
In the hydrogen disk target,
the protons are accelerated more efficiently than in the carbon case,
because hydrogen is much ``lighter'' than carbon.

Because I used the same laser pulse in all simulations,
the numbers of electrons pushed out from the target
are estimated to be almost the same in carbon and hydrogen disk cases
at the initial simulation time.
Therefore, the target surface charge is almost the same in the two cases,
and the proton energy by the charged disk electric field can be estimated
using the same considerations as in Section \ref{sim-b},
yielding $35$ MeV.
Therefore, we can estimate that the proton energy
by RPDA and Coulomb explosion is $\approx$165 MeV.
We can say that the RPDA and Coulomb explosion effect is much stronger in the
hydrogen disk case compared with the carbon case.

We see that
protons are distributed in different areas based on each energy level
(see Fig. \ref{fig:fig-h1}).
Moreover,
high-energy protons are distributed on the $+x$ side edge of the proton cloud
and are moving in the $+x$ direction.
Therefore,
we could select only high-energy protons by using a pinhole with a shutter.
The shutter must have high enough accuracy and the ability to shield unwanted
particles and radiation.
The shutter speed must be very fast near the target.
The accuracy becomes coarser
if the shutter position moves to a position farther from the target,
because the distance between the high-energy protons and the low-energy protons
grows in time.
It may be difficult to construct a shutter that satisfies both the timing
accuracy and shielding ability,
however, we could get a similar result by using a magnet and a slit.
The path of a proton is changed for each energy level by a magnetic field,
and high-energy protons can be taken out by passing through a slit.
I show the results using the previous method.
\begin{figure}[tbp]
\includegraphics[clip,width=7.0cm,bb=0 0 495 487]{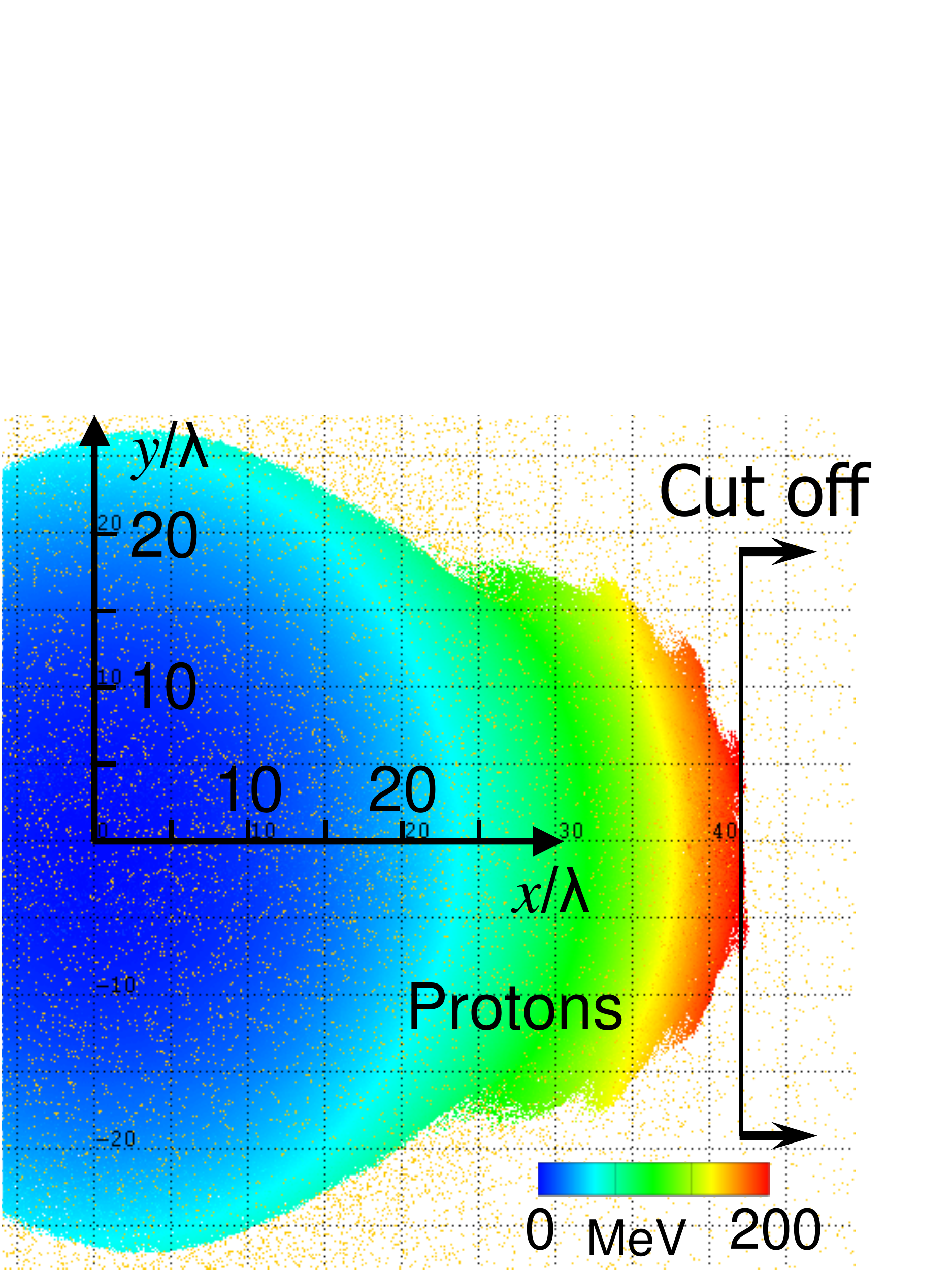}
\caption{
Distributions of protons at $t=100\times 2\pi/\omega$
and the selected area of the proton bunch.
}
\label{fig:fig-h4}
\end{figure}
\begin{figure}[tbp]
\includegraphics[clip,width=8.0cm,bb=20 5 527 440]{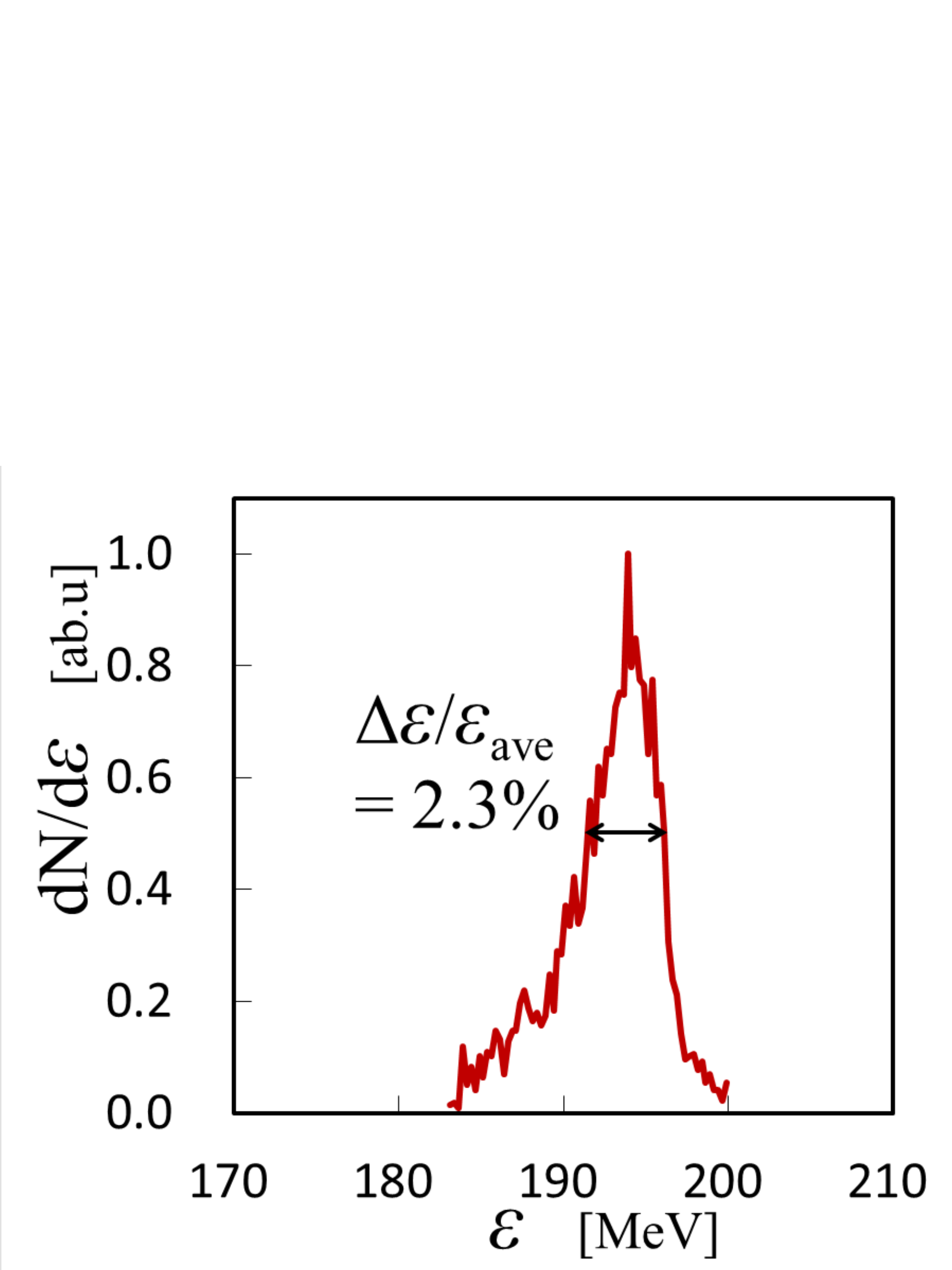}
\caption{
Proton energy spectrum obtained by cutting off the proton bunch
at $t=100\times 2\pi/\omega$, normalized by the maximum.
}
\label{fig:fig-h5}
\end{figure}
The cutoff position is shown in Fig. \ref{fig:fig-h4}.
Figure \ref{fig:fig-h5} shows energy spectrum of cutoff protons
at $t=100\times 2\pi/\omega$.
We obtain a proton beam with
a maximum energy of $\mathcal{E}_\mathrm{max}=200$ MeV and
an average energy of $\mathcal{E}_\mathrm{ave}=193$ MeV with
an energy spread of $\Delta\mathcal{E}/\mathcal{E}_\mathrm{ave}=2.3\%$
and a particle number of $2.6\times10^7$.
This proton beam has high enough energy and quality for some applications
(e.g., in medical applications).

\section{CONCLUSIONS}
Proton acceleration driven by a laser pulse irradiating a disk target
is investigated with the help of 3D PIC simulations.
I have found
higher energy protons are obtained by using
``light'' materials for the target.
As seen in simulations, for these materials
a strongly inhomogeneous expansion of the disk target occurs
owing to the Coulomb explosion, which plays an important role,
and RPDA has a strong effect.
The time-varying electric potential of the
inhomogeneous expanding ion cloud
and the movement of the ion cloud for the protons
efficiently accelerate protons.
The proton beam energy can be substantially increased
by using a ``light'' material for the target.
In our simulations, the laser intensity and energy,
$I_0=5\times 10^{21}$ W/cm$^{2}$ and $\mathcal{E}_{las}=18$ J,
are not enough to reach the RPDA regime in full scale, but
the RPDA regime has a big effect even at this laser power level.
Although I show simulation results by using a simple hydrogen disk,
it may be difficult to fabricate such a hydrogen disk;
a CH$_n$ foil target with a high $n$ value should be a good substitute.
The laser parameters used in this paper---intensity, power, energy,
and spot size---are ones already existing in current laser systems.
Therefore, we should be able to generate $200$-MeV protons now.

\section*{ACKNOWLEDGMENTS}
I thank P. Bolton, S. V. Bulanov, T. Esirkepov, M. Kando, J. Koga, K. Kondo,
and M. Yamagiwa for useful discussions.
The computations were performed using the PRIMERGY BX900 supercomputer
at JAEA Tokai.
This work was partially supported by
the Ministry of Education, Culture, Sports, Science and Technology
Grant-in-Aid for Scientific Research (C) No. 23540584.


\begin{thebibliography}{99}

\bibitem{SBK}
S. V. Bulanov and V. S. Khoroshkov, Plasma Phys. Rep. \textbf{28}, 453 (2002);
S. V. Bulanov, T. Zh. Esirkepov, V. S. Khoroshkov, A. V. Kuznetsov,
and F. Pegoraro, Phys. Lett. A \textbf{299}, 240 (2002).

\bibitem{MVL}
E. Fourkal, I. Velchev, J. Fan, W. Luo, and C. Ma, Med. Phys. \textbf{34} 577 (2007).

\bibitem{ROT}
M. Roth, T. E. Cowan, M. H. Key, S. P. Hatchett, C. Brown, W. Fountain,
J. Johnson, D. M. Pennington, R. A. Snavely, S. C. Wilks, K. Yasuike, H. Ruhl,
F. Pegoraro, S. V. Bulanov, E. M. Campbell, M. D. Perry, and H. Powell,
Phys. Rev. Lett. \textbf{86}, 436 (2001).

\bibitem{BRM}
V. Yu. Bychenkov, W. Rozmus, A. Maksimchuk, D. Umstadter, and C. E. Capjack,
Plasma Phys. Rep. \textbf{27}, 1017 (2001).

\bibitem{ATH}
S. Atzeni, M. Temporal, and J. J. Honrubia,
Nucl. Fusion \textbf{42}, L1 (2002).

\bibitem{ESI1}
T. Esirkepov, M. Borghesi, S. V. Bulanov, G. Mourou, and T. Tajima,
Phys. Rev. Lett. \textbf{92}, 175003 (2004).

\bibitem{BEE}
S. V. Bulanov, E. Yu. Echkina, T. Zh. Esirkepov, I. N. Inovenkov, M. Kando,
F. Pegoraro, and G. Korn,
Phys. Rev. Lett. \textbf{104}, 135003 (2010).

\bibitem{BWP}
J. Badziak, E. Woryna, P. Parys, K. Yu. Platonov, S. Jablo\'{n}ski,
L. Ry\'{c}, A. B. Vankov, and J. Woowski,
Phys. Rev. Lett. \textbf{87}, 215001 (2001).

\bibitem{FVM}
E. Fourkal, I. Velchev, and C.-M. Ma,
Phys. Rev. E \textbf{71}, 036412 (2005).

\bibitem{Toncian}
T. Toncian, M. Borghesi, J. Fuchs, E. d'Humi\`{e}res, P. Antici, P. Audebert,
E. Brambrink, C. A. Cecchetti, A. Pipahl, L. Romagnani, and O. Willi,
Science {\bf 312}, 410 (2006).

\bibitem{HAC}
B. M. Hegelich, B. J. Albright, J. Cobble, K. Flippo, S. Letzring, M. Paffett,
H. Ruhl, J. Schreiber, R. K. Schulze, and J. C. Fern\'{a}ndez,
Nature (London) \textbf{439}, 441 (2006).

\bibitem{YAH}
L. Yin, B. J. Albright, B. M. Hegelich, K. J. Bowers, K. A. Flippo,
T. J. T. Kwan, and J. C. Fern\'{a}ndez,
Phys. Plasmas \textbf{14}, 056706 (2007).

\bibitem{PRK}
A. P. L Robinson, A. R. Bell, and R. J. Kingham,
Phys. Rev. Lett. \textbf{96}, 035005 (2006).

\bibitem{PPM}
F. Peano, F. Peinetti, R. Mulas, G. Coppa, and L. O. Silva,
Phys. Rev. Lett. \textbf{96}, 175002 (2006).

\bibitem{HSM}
M. Hohenberger, D. R. Symes, K. W. Madison, A. Sumeruk, G. Dyer, A. Edens,
W. Grigsby, G. Hays, M. Teichmann, and T. Ditmire,
Phys. Rev. Lett. \textbf{95} 195003 (2005).

\bibitem{DL}
T. Esirkepov, S. V. Bulanov, K. Nishihara, T. Tajima, F. Pegoraro,
V. S. Khoroshkov, K. Mima, H. Daido, Y. Kato, Y. Kitagawa, K. Nagai,
and S. Sakabe,
Phys. Rev. Lett. \textbf{89}, 175003 (2002).

\bibitem{SPJ}
H. Schwoerer, S. Pfotenhauer, O. J\"{a}ckel, K.-U. Amthor, B. Liesfeld,
W. Ziegler, R. Sauerbrey, K. W. D. Ledingham, and T. Esirkepov,
Nature (London) \textbf{439}, 445 (2006).

\bibitem{MEBKY}
T. Morita, T. Zh. Esirkepov, S. V. Bulanov, J. Koga, and M. Yamagiwa,
Phys. Rev. Lett. \textbf{100}, 145001 (2008).

\bibitem{MEBKY2}
T. Morita, S.V. Bulanov, T. Zh. Esirkepov, J. Koga, and M. Yamagiwa,
Phys. Plasmas \textbf{16}, 033111 (2009).

\bibitem{CBL}
C. K. Birdsall and A. B. Langdon, \textit{Plasma Physics via Computer Simulation}
(McGraw-Hill, New York, 1985).

\bibitem{HSB}
K. Harres, M. Schollmeier, E. Brambrink, P. Audebert, A. Bla\v{z}evi\'{c},
K. Flippo, D. C. Gautier, M. GeiBel, B. M. Hegelich, F. N\"urnberg,
J. Schreiber, H. Wahl, and M. Roth,
Rev. Sci. Instrum. \textbf{79}, 093306 (2008).

\bibitem{HKM}
A. Henig, D. Kiefer, K. Markey, D. C. Gautier, K. A. Flippo, S. Letzring,
R. P. Johnson, T. Shimada, L. Yin, B. J. Albright, K. J. Bowers,
J. C. Fern\'andez, S. G. Rykovanov, H.-C. Wu, M. Zepf, D. Jung,
V. Kh. Liechtenstein, J. Schreiber, D. Habs, and B. M. Hegelich,
Phys. Rev. Lett. \textbf{103}, 045002 (2009).

\bibitem{JHK}
D. Jung, R. H\"orlein, D. Kiefer, S. Letzring, D. C. Gautier, U. Schramm,
C. H\"ubsch, R. \"Ohm, B. J. Albright, J. C. Fernandez, D. Habs,
and B. M. Hegelich,
Rev. Sci. Instrum. \textbf{82}, 013306 (2011).

\bibitem{SNAV}
R. A. Snavely, M. H. Key, S. P. Hatchett, T. E. Cowan, M. Roth, T. W. Phillips,
M. A. Stoyer, E. A. Henry, T. C. Sangster, M. S. Singh, S. C. Wilks,
A. MacKinnon, A. Offenberger, D. M. Pennington, K. Yasuike, A. B. Langdon,
B. F. Lasinski, J. Johnson, M. D. Perry, and E. M. Campbell,
Phys. Rev. Lett. \textbf{85}, 2945 (2000).

\bibitem{CLAR}
E. L. Clark, K. Krushelnick, J. R. Davies, M. Zepf, M. Tatarakis, F. N. Beg,
A. Machacek, P. A. Norreys, M. I. K. Santala, I. Watts, and A. E. Dangor,
Phys. Rev. Lett. \textbf{84}, 6703 (2000).

\bibitem{MBEKK}
T. Morita, S. V. Bulanov, T. Zh. Esirkepov, J. Koga, and M. Kando,
J. Phys. Soc. Jpn. \textbf{81}, 024501 (2012).

\end{thebibliography}
\end{document}